\documentclass[aps,prd,twocolumn,showpacs]{revtex4} 

\usepackage{graphicx}

\begin{document}

\title{Simultaneous Flavor Transformation of Neutrinos and Antineutrinos
with Dominant Potentials from Neutrino-Neutrino Forward Scattering}
\author{George M.\ Fuller} 
\affiliation{Department of Physics, University of California, San Diego, 
La Jolla, CA 92093-0319}
\author{Yong-Zhong Qian} 
\affiliation{School of Physics and Astronomy, University of Minnesota,
Minneapolis, MN 55455}
\date{\today}

\begin{abstract}
In astrophysical environments with intense neutrino fluxes,
neutrino-neutrino forward scattering contributes both diagonal and
off-diagonal potentials to the flavor-basis Hamiltonian that governs
neutrino flavor evolution. We examine a special case where adiabatic
flavor evolution can produce an off-diagonal potential from
neutrino-neutrino forward scattering that dominates over both the 
corresponding diagonal term and the potential from neutrino-matter 
forward scattering. In this case, we find a solution that, unlike 
the ordinary Mikeyhev-Smirnov-Wolfenstein scenario, has both 
neutrinos and antineutrinos maximally mixed in medium over appreciable 
ranges of neutrino and antineutrino energy. Employing the measured 
solar and atmospheric neutrino mass-squared differences,
we identify the conditions on neutrino fluxes
that are required for this solution to exist deep in the supernova 
environment, where it could affect the neutrino signal, 
heavy-element nucleosynthesis, and even the revival of the supernova 
shock. We speculate on how this solution might or might not be attained 
in realistic supernova evolution. Though this solution is ephemeral in 
time and/or space in supernovae, it may signal the onset of subsequent 
appreciable flavor mixing for both neutrinos and antineutrinos.
A similar solution may also exist in an early universe with
significant net neutrino-lepton numbers.
\end{abstract}
\pacs{97.60.Gb,14.60.Pq}
\maketitle

%%%%%%%%%%%%%%%%%%%%%%%%%%%%%%%%%%%%%%%%%%%%%%%%%%%%%%%%%%%%%%%%%%%%
\section{Introduction\label{sec-1}}
%%%%%%%%%%%%%%%%%%%%%%%%%%%%%%%%%%%%%%%%%%%%%%%%%%%%%%%%%%%%%%%%%%%%
In this paper we study the problem of coherent nonlinear
flavor evolution of active neutrinos in environments where
neutrino-neutrino forward scattering makes a significant
contribution to the effective neutrino mass in medium.
In particular, we examine a special case where
the off-diagonal potential from neutrino-neutrino 
forward scattering becomes the dominant term in the 
flavor-basis Hamiltonian that governs neutrino flavor evolution. 

In both post-core-bounce supernovae and an early universe with
net lepton numbers, local net neutrino number densities can exceed 
electron and baryon number densities. These large neutrino number 
densities or fluxes, sometimes referred to as a \lq\lq neutrino 
background,\rq\rq\ require that the usual Mikheyev-Smirnov-Wolfenstein
(MSW) formalism \cite{MSW} for calculating the evolution of neutrino 
flavors be modified to include the effects of neutrino-neutrino
forward scattering. Though the resulting problem of neutrino flavor 
evolution can be complicated, we can identify a key parameter
governing the relevant physics: the ratio of the product of neutrino 
energy and the off-diagonal potential from neutrino-neutrino forward 
scattering relative to the difference of the squares of the vacuum 
neutrino mass eigenvalues. If this ratio becomes very large as a
result of adiabatic neutrino flavor evolution, we can even find
an interesting solution, which has neutrinos and antineutrinos
simultaneously maximally mixed in medium. This is different 
from the usual MSW case where, at a given location or time, neutrino 
flavor mixing in medium is maximal for a specific range of neutrino 
energies (this range is narrow for small vacuum mixing angles), while 
antineutrino mixing is suppressed; or {\it vice versa.} As we discuss 
below, the off-diagonal potential from neutrino-neutrino forward 
scattering plays a unique role: it can alter neutrino
flavor evolution into a form that is utterly 
unlike the MSW case. 

In the presence of a significant neutrino background, neutrino flavor 
histories can be followed by solving a mean-field Schr\"odinger-like 
equation in the modified-MSW format. This is a
long-standing and vexing problem. In fact, it has defied
general and complete solution for the supernova
environment, even with sophisticated numerical treatments.
The existence and importance of the flavor-diagonal potential
from neutrino-neutrino forward scattering and how it might modify 
MSW-like neutrino flavor evolution in supernovae were pointed out 
early on \cite{FMWS} (see Ref.~\cite{RS} for a subsequent formal treatment).
However, the existence of the corresponding off-diagonal potential 
was established only later \cite{Pant}. This latter discovery may prove 
to be a watershed event in supernova neutrino physics. 

There have been several attempts to elucidate how flavor-diagonal
and/or off-diagonal potentials from neutrino-neutrino forward 
scattering can affect active-active neutrino flavor transformation
in the post-core-bounce supernova regime, especially
as regards shock re-heating \cite{FM,QF95,Mez}
and $r$-process nucleosynthesis \cite{QF95,Qian,Horow,Pastor}. The 
rationale for these studies was that the energy spectra and/or
the fluxes of the various neutrino flavors could differ on emergence
from the neutron star surface or neutrino sphere, and therefore,
flavor inter-conversion above this surface could alter these spectra 
and/or fluxes to change supernova dynamics and nucleosynthesis or the 
neutrino signal in a detector. If the energy distribution functions 
and the associated net energy luminosities are
the same for all flavors of neutrinos and antineutrinos,
then, obviously, flavor transformation will have no effect. However, 
on account of core de-leptonization
and concomitant changes in composition, size, and equation of state,
this is unlikely to be the case over the entire post-core-bounce
period of $\sim 20\,{\rm s}$ during which neutrino
fluxes are appreciable.

If at any epoch in the supernova environment there develops a hierarchy 
of average neutrino energies or luminosities among
the different neutrino flavors, then flavor conversion could alter 
the rates of electron neutrino and antineutrino
capture on free nucleons:
\begin{eqnarray}
\label{rates1}
\nu_e+n  & \rightarrow & p+e^-,
\\
\label{rates2}
\bar\nu_e+p & \rightarrow & n+e^+.
\end{eqnarray}
These are the processes principally responsible
for depositing energy in the material behind the shock after core bounce.
Therefore, altering their rates by, for example, swapping flavor
labels between possibly less energetic electron neutrinos and 
more energetic mu and/or tau neutrinos could significantly affect
the prospects for a supernova explosion \cite{FM}. Also, the competition
between these processes and their reverse reactions sets 
the neutron-to-proton ratio in neutrino-heated material \cite{Qian}.
In turn, this ratio is sometimes a crucial parameter for
$r$-process \cite{Qian,QF95} and other heavy-element nucleosynthesis 
\cite{Hoffman} associated with slow neutrino-heated outflows.

Most of the studies cited above posited the existence of
neutrino mass-squared differences $\ge 0.2\,{\rm eV}^2$.
This was required for normal MSW resonances to occur
in the high-density regions most relevant for supernova shock re-heating 
and $r$-process nucleosynthesis. These regions lie above but relatively
close to the neutron star, generally within a few hundred kilometers. 
Without the hypothesized high neutrino mass-squared differences,
conventional MSW evolution in these regions
would not result in any significant neutrino flavor conversion.

Although we do not
know the absolute vacuum neutrino mass eigenvalues,
$m_1$, $m_2$, and $m_3$, the
two independent differences
of their squares are now measured to be
$\delta m^2 \approx 7\times{10}^{-5}$ and
$3\times{10}^{-3}\,{\rm eV}^2$
by observations of solar and atmospheric neutrinos, respectively.
The lower $\delta m^2$ has also been measured directly by the 
KamLAND reactor experiment. (See Ref.~\cite{nuref} 
for a review of neutrino properties.) 
As these $\delta m^2$ values are certainly small compared with
the scale previously believed to be most relevant
for supernovae, one may tend to conclude that they have no 
consequence for supernova dynamics and nucleosynthesis. However,  
as we will discuss below, neutrino background effects
could alter this conclusion dramatically. 

Similar to the supernova case, when there are net lepton numbers residing
in the neutrino sector in the early universe, neutrino flavor 
conversion can be important in, for example, setting the neutron-to-proton
ratio and, hence, the $^4${He} abundance yield in primordial
nucleosynthesis. It was recognized in Ref. \cite{SMF} that the flavor-diagonal 
potential from neutrino-neutrino forward scattering affects neutrino 
propagation in the coherent limit of the problem. However, for reasons 
that will become clear, a complete treatment of 
active-active neutrino flavor conversion in the early universe
requires a coupled calculation including both flavor-diagonal and
off-diagonal potentials from neutrino-neutrino forward scattering
as well as general inelastic neutrino scattering 
\cite{FV,abb,aba}. In fact, the seminal numerical work
in Refs. \cite{abb,aba} shows that the measured neutrino mass-squared
differences and mixing angles result in an
\lq\lq evening up\rq\rq\ of the initially disparate
lepton numbers residing in neutrinos
of different flavors. The flavor oscillations
in these calculations exhibit \lq\lq synchronization\rq\rq\
in time and space and can correspond to near-maximal flavor mixing for 
neutrinos and antineutrinos. We will argue below that this numerical
result is closely related to our solution for the special case where
the off-diagonal potential from neutrino-neutrino forward scattering
dominates.

In Sec. II we will outline how neutrino flavor
transformation proceeds in the coherent limit
when neutrino backgrounds are non-negligible.
In Sec. III we discuss the particular limit of domination by
the off-diagonal potential from neutrino-neutrino forward scattering
and the corresponding solution.
We also speculate under what conditions and to what
extent this solution could be attained. 
Similar issues for a lepton-degenerate early universe are discussed in Sec. IV.
Conclusions are given in Sec. V.

\section{Coherent Flavor Evolution with Neutrino Backgrounds\label{sec-2}}

Here we give a brief synopsis of coherent neutrino
flavor amplitude development in the supernova and early
universe environments. In the supernova core and 
the dense environment immediately above it, and in
the early universe prior to weak decoupling, non-forward neutrino
scattering can result in neutrino flavor conversion through
de-coherence. We will ignore this in what follows and instead concentrate
on the purely coherent evolution of the neutrino fields. It should
always be kept in mind, however, that our considerations may need to be
modified at high density or in high neutrino flux regimes.

Even in the purely coherent limit, following the effects of
neutrino-neutrino forward scattering in the most general case
is daunting in scope.
Part of the difficulty in following
neutrino flavor evolution in the supernova environment is geometric:
flavor evolution histories on different neutrino trajectories are
coupled. This is because two neutrino states 
will experience quantum entanglement to the future of
a forward scattering event occurring at the intersection of their world lines.
In light of this entanglement there has been considerable speculation about 
whether neutrino flavor evolution can be 
modeled adequately by a mean-field treatment with Schr\"odinger
equations \cite{AF03,BRS03}. The mean field in this case
is the potential seen by a neutrino by
virtue of forward scattering on particles in the environment that carry 
weak charge. 
Here we will follow the conclusions
of Ref. \cite{AF03} and take the mean-field treatment as sufficient in
a statistical sense. This seems reasonable for the early universe and 
supernova environments
because a statistically large number of neutrino
scattering events and entanglements occur in these places. 

Further complicating the supernova problem is the
non-isotropic nature of the neutrino fields above the neutrino sphere.
Neutrinos traveling along trajectories nearly tangential
to the neutrino sphere may have quite different
flavor amplitude histories from those moving radially
or near radially. For now we will ignore this feature of 
flavor development and instead
approximate all neutrinos as evolving the way
radially propagating neutrinos evolve. This approximation
is {\it not} a good one (it has been made in all previous numerical work 
\cite{QF95, Pastor}),
but it will suffice in our analytic arguments here.

\subsection{Overview\label{sec-2a}}

In vacuum, the flavor (weak interaction) eigenstates of neutrinos are
related to the mass (energy) eigenstates by a unitary transformation 
$U_m$:
\begin{equation}
\left(\begin{array}{c}
|\nu_{e}\rangle\\
|\nu_{\mu}\rangle\\
|\nu_{\tau}\rangle\end{array}\right)
 =U_m\left(\begin{array}{c}
|\nu_1\rangle\\
|\nu_2\rangle\\
|\nu_{3}\rangle\end{array}\right),
\label{unitary}
\end{equation}
where the mass eigenstates $\vert \nu_1\rangle$, 
$\vert \nu_2\rangle$, and $\vert \nu_2\rangle$ correspond to the
vacuum mass eigenvalues, $m_1$, $m_2$, and $m_3$, respectively.
The unitary transformation $U_m$ can be written in terms of
a sequence of rotations,
\begin{equation}
U_m=U_{23}U_{13}U_{12}.
\label{unitaryrot}
\end{equation}
A convenient representation for these rotations is
\begin{equation}
\begin{array}{ccc}
U_{23} & \equiv & \left(\begin{array}{ccc}
1 & 0 & 0\\
0 & \cos\theta_{23} & \sin\theta_{23}\\
0 & -\sin\theta_{23} & \cos\theta_{23}\end{array}\right),\\
U_{13} & \equiv & \left(\begin{array}{ccc}
\cos\theta_{13} & 0 &e^{i\delta} \sin\theta_{13}\\
0 & 1 & 0\\
-e^{-i\delta}\sin\theta_{13} & 0 & \cos\theta_{13}\end{array}\right),\\
U_{12} & \equiv & \left(\begin{array}{ccc}
\cos\theta_{12} & \sin\theta_{12} & 0\\
-\sin\theta_{12} & \cos\theta_{12} & 0\\
0 & 0 & 1\end{array}\right).
\label{rots}
\end{array}
\end{equation}
In the above representation, 
the mixing angles, $\theta_{12}$ and $\theta_{23}$, have been measured
by observations of solar and atmospheric neutrinos and related experiments.
In particular, the best fit for the vacuum mixing of the mu and tau 
neutrinos is very near maximal, which gives $\theta_{23}\approx \pi/4$. 
However, the mixing angle, $\theta_{13}$, and the CP-violating phase, 
$\delta$, have not been measured yet.

Here we consider a neutrino mixing scenario where $m_3>m_2>m_1$ with
$\delta m_{12}^2\equiv m_2^2-m_1^2\approx 7\times{10}^{-5}\,{\rm eV}^2$ and
$\delta m_{13}^2\equiv m_3^2-m_1^2\approx 3\times{10}^{-3}\,{\rm eV}^2$,
$\theta_{23}=\pi/4$, and $\delta=0$. With the definitions
\begin{eqnarray}
\vert \nu_\mu^\ast\rangle & \equiv & 
{{\vert \nu_\mu\rangle -\vert \nu_\tau\rangle}\over{\sqrt{2}}},\\
\vert \nu_\tau^\ast\rangle & \equiv & 
{{\vert \nu_\mu\rangle +\vert \nu_\tau\rangle}\over{\sqrt{2}}},
\label{taustar}
\end{eqnarray}
it is straightforward to show that
\begin{equation}
\left(\begin{array}{c}
|\nu_{e}\rangle\\
|\nu_{\mu}^*\rangle\\
|\nu_{\tau}^*\rangle\end{array}\right)
 =\left(\begin{array}{ccc}
c_{12}c_{13}&s_{12}c_{13}&s_{13}\\
-s_{12}&c_{12}&0\\
-c_{12}s_{13}&-s_{12}s_{13}&c_{13}\end{array}\right)
\left(\begin{array}{c}
|\nu_1\rangle\\
|\nu_2\rangle\\
|\nu_{3}\rangle\end{array}\right),
\label{3nu}
\end{equation}
where for example, $c_{12}\equiv\cos\theta_{12}$ and
$s_{12}\equiv\sin\theta_{12}$. The states $\vert\nu_\mu^\ast\rangle$ and
$\vert\nu_\tau^\ast\rangle$ are still useful in medium. This is because 
in the supernova medium the mu and tau neutrinos have very nearly the 
same interactions, so that matter effects on mixing and effective mass 
for these species are nearly identical (likewise for the mu and tau 
antineutrinos). This will also be true for the early universe if
the net muon and tau lepton numbers are
identical. For the sake of our arguments here, we will take the symmetry 
between mu and tau neutrinos and that between their antiparticles to be 
rigorously true so that $\vert\nu_\mu^\ast\rangle$ and
$\vert\nu_\tau^\ast\rangle$ are effective flavor eigenstates in medium.
Since $\delta m_{13}^2\gg\delta m_{12}^2$,
the regions of neutrino flavor mixing governed by these parameters should
be well separated. As neutrinos propagate outward from the neutrino
sphere at high density in supernovae, 
$\delta m_{13}^2\approx 3\times{10}^{-3}\,{\rm eV}^2$
becomes relevant first. We will focus on neutrino flavor mixing with
this parameter, for which $\nu_\mu^\ast$ is effectively decoupled
[see Eq.~(\ref{3nu})]
and we only need consider mixing of $\nu_e$ and $\nu_\tau^\ast$
\cite{BF,CFQ}. Thus, the general problem of 3$\nu$ mixing in medium is 
reduced to one of 2$\nu$ mixing in our scenario.

With the above simplification we can hereafter follow the notation of
Ref. \cite{QF95}. In particular, we now simply refer to 
$\vert \nu_\tau^\ast\rangle$ as $\vert \nu_\tau\rangle$ and
write the effective $2\nu$ unitary transformation in vacuum as
\begin{eqnarray}
\vert \nu_e\rangle & = & \cos\theta \vert \nu_1\rangle + 
\sin\theta \vert\nu_2\rangle
\label{vac2x21}
\\
\vert \nu_\tau\rangle & = & -\sin\theta \vert \nu_1\rangle + 
\cos\theta \vert\nu_2\rangle,
\label{vac2x22}
\end{eqnarray}
where $\vert \nu_1\rangle$ and $\vert\nu_2\rangle$ refer generically to 
the light and heavy mass eigenstates, respectively, and $\theta$ is the 
effective $2\nu$ vacuum mixing angle. The relevant vacuum 
mass-squared difference is $\delta m^2\approx 3\times{10}^{-3}\,{\rm eV}^2$.
The corresponding effective vacuum mixing angle is $\theta\sim\theta_{13}$, 
with the current reactor experiment limit 
being $\sin^22\theta_{13} < 0.1$ (see e.g., Ref.~\cite{nuref}). 

Consider a neutrino of initial flavor $\alpha=e$ or $\tau$.
As it propagates outward from the neutrino sphere in supernovae,
the evolution of its state, the ket $\vert \Psi_{\nu_\alpha}(t)\rangle$,
can be described as
\begin{equation}
\vert\Psi_{\nu_\alpha}(t)\rangle = a_{e\alpha}(t)\vert\nu_e\rangle+
a_{\tau\alpha}(t)\vert\nu_\tau\rangle,
\end{equation}
where $a_{e\alpha}(t)$ or $a_{\tau\alpha}(t)$ is the amplitude for
the neutrino to be a $\nu_e$ or $\nu_\tau$, respectively, at time $t$. (Note
that $t$ could be any Affine parameter such as radius along the neutrino's world line.)
Alternatively, the evolution of $\vert \Psi_{\nu_\alpha}(t)\rangle$ can 
be described as
\begin{equation}
\vert\Psi_{\nu_\alpha}(t)\rangle = a_{1\alpha}(t)\vert\nu_1(t)\rangle+
a_{2\alpha}(t)\vert\nu_2(t)\rangle,
\label{single}
\end{equation}
where $\vert\nu_1(t)\rangle$ and $\vert\nu_2(t)\rangle$ are
the instantaneous mass (energy) eigenstates in medium, and
$a_{1\alpha}(t)$ and $a_{2\alpha}(t)$ are the corresponding amplitudes.
The flavor eigenstates
are related to $\vert\nu_1(t)\rangle$ and $\vert\nu_2(t)\rangle$ as
\begin{eqnarray}
\vert \nu_e\rangle & = & \cos\theta_M(t) \vert \nu_1(t)\rangle + 
\sin\theta_M(t) \vert\nu_2(t)\rangle
\\
\vert \nu_\tau\rangle & = & -\sin\theta_M(t) \vert \nu_1(t)\rangle + 
\cos\theta_M(t) \vert\nu_2(t)\rangle,
\label{medium2x2}
\end{eqnarray}
where $\theta_M(t)$ is the effective $2\nu$ mixing angle in medium
at time $t$. In matrix form, the ket $\vert \Psi_{\nu_\alpha}(t)\rangle$
can be represented by
\begin{equation}
\Psi_f\equiv\left[\begin{array}{c}
a_{e\alpha}(t)\\
a_{\tau\alpha}(t)\end{array}\right]
\end{equation}
in the flavor basis and by
\begin{equation}
 \Psi_M\equiv\left[\begin{array}{c}
a_{1\alpha}(t)\\
a_{2\alpha}(t)
\end{array}\right]
\label{medium}
\end{equation}
in the energy basis. In analogous fashion we will employ a $2\bar\nu$ scheme 
to follow separately the flavor evolution of the antineutrino sector.

\subsection{Characterizing Neutrino Densities
\label{sec:nuden}}

The single neutrino density operator at time $t$ projected into the energy 
basis is
\begin{widetext} 
\begin{eqnarray}
\vert\Psi_{\nu_\alpha}(t)\rangle\langle\Psi_{\nu_\alpha}(t)\vert   
& = & {\vert a_{1\alpha}(t)\vert}^2\vert\nu_1(t)\rangle\langle\nu_1(t)\vert
+{\vert a_{2\alpha}(t)\vert}^2\vert\nu_2(t)\rangle\langle \nu_2(t)\vert
\nonumber
\\
& + &{a_{1\alpha}(t)a_{2\alpha}^\ast}(t)\vert\nu_1(t)\rangle\langle\nu_2(t)\vert 
+{a_{1\alpha}^\ast(t)a_{2\alpha}(t)}\vert\nu_2(t)\rangle\langle \nu_1(t)\vert. 
\label{singleop}
\end{eqnarray}
\end{widetext}
The second line of Eq.~(\ref{singleop}) contains cross terms which in general 
have complex coefficients. However, these cross terms vanish in the limit 
where neutrino flavor evolution is adiabatic. This is because a neutrino 
evolving adiabatically is always in a single energy state.
For example, in this limit we might have $\vert a_{1\alpha}(t)\vert=1$,
which would imply that $\vert a_{2\alpha}(t)\vert=0$ due to the
normalization condition
\begin{equation}
\langle\Psi_\alpha(t)\vert\Psi_\alpha(t)\rangle=\vert a_{1\alpha}(t)\vert^2+
\vert a_{2\alpha}(t)\vert^2=1.
\label{norm}
\end{equation}

The density operator for the neutrinos or antineutrinos with momentum centered 
around ${\bf p}$ in a pencil of neutrino or antineutrino momenta and directions 
$d^3{\bf p}$ can be defined as in Ref. \cite{QF95}:
\begin{eqnarray}
{\hat{\rho}}_{\bf p}\left( t\right)d^3{\bf p} & \equiv & 
\sum_{\alpha} dn_{\nu_\alpha}\vert\Psi_{\nu_\alpha}(t)\rangle
\langle\Psi_{\nu_\alpha}(t)\vert,
\label{deopn}
\\
{\hat{\bar\rho}}_{\bf p}\left( t\right)d^3{\bf p} & \equiv & 
\sum_{\alpha} dn_{\bar\nu_\alpha}\vert\Psi_{\bar\nu_\alpha}(t)\rangle
\langle\Psi_{\bar\nu_\alpha}(t)\vert.
\label{denopan}
\end{eqnarray}
Note that the traces of these operators over neutrino flavor do not give 
unity but rather the total number density of neutrinos or antineutrinos of all kinds 
in the pencil.

We assume that neutrinos and antineutrinos of all flavors are emitted 
from the same sharp neutrino sphere of radius $R_\nu$ in supernovae. 
(This is not a particularly good approximation for neutrinos very near the 
neutron star surface, but it will suffice for our arguments.) At a radius
$r>R_\nu$, the neutrino sphere subtends a solid angle of
\begin{equation}
\Delta\Omega_\nu(r)=2\pi\left(1-\sqrt{1-{R_\nu^2/r^2}}\right).
\label{nusolid}
\end{equation}
Within this solid angle,
the number density of $\nu_\alpha$ in a pencil of directions and momenta is 
\begin{equation}
dn_{\nu_\alpha}={{L_{\nu_\alpha}}\over{\pi R_\nu^2}} 
{{1}\over{\langle E_{\nu_\alpha}\rangle}}  
{\left( {{d\Omega_\nu}\over{4\pi}} \right)} 
f_{\nu_\alpha}\left(E_\nu\right) dE_\nu,
\label{normdist}
\end{equation}
where $L_{\nu_\alpha}$ is the energy luminosity of $\nu_\alpha$,
$d\Omega_\nu$ is the pencil of
directions, $E_\nu$ is the neutrino energy, $f_{\nu_\alpha}(E_\nu)$
is the normalized energy distribution function for $\nu_\alpha$, and 
$\langle E_{\nu_\alpha}\rangle$ is the corresponding average $\nu_\alpha$ 
energy. Here and in the rest of this paper we assume that neutrinos have 
relativistic kinematics and employ natural units where $\hbar=c=1$. 
The function $f_{\nu_\alpha}(E_\nu)$ can be fitted to the results 
from supernova neutrino transport calculations and is commonly taken to be 
of the form
\begin{equation}
f_{\nu_\alpha}(E_\nu) = {{1}\over{T_{\nu_\alpha}^3 F_2(\eta_{\nu_\alpha})}} 
{{E_\nu^2}\over{e^{E_\nu/T_{\nu_\alpha}-\eta_{\nu_\alpha}}+1}},
\label{disn}
\end{equation}
where $T_{\nu_\alpha}$ and $\eta_{\nu_\alpha}$ are two fitting
parameters and
$F_2(\eta_{\nu_\alpha})$ is the Fermi integral of order 2 and argument
$\eta_{\nu_\alpha}$. The Fermi integral of order $k$ and argument $\eta$ is
defined as
\begin{equation}
F_k\left(\eta\right)\equiv \int_0^\infty{{x^kdx}\over{e^{x-\eta}+1}}.
\label{fermint}
\end{equation}
In terms of these integrals, the average $\nu_\alpha$ energy is 
\begin{equation}
\langle E_{\nu_\alpha}\rangle\equiv\int_0^\infty{E_\nu 
f_{\nu_\alpha}\left(E_\nu\right)dE_\nu }=T_{\nu_\alpha} 
{{F_3\left(\eta_{\nu_\alpha}\right)}\over{F_2\left(\eta_{\nu_\alpha}\right)}}.
\label{average}
\end{equation}

We assume that all neutrino species have thermal, Fermi-Dirac energy 
distribution functions in the early universe, so the number density of 
$\nu_\alpha$ in a pencil of directions and momenta is 
\begin{equation}
dn_{\nu_\alpha}={{1}\over{2\pi^2}} {\left( {{d\Omega_\nu}\over{4\pi}} \right)} 
{{E_\nu^2 dE_\nu}\over{e^{E_\nu/T_{\nu_\alpha}-\eta_{\nu_\alpha}}+1}}. 
\label{dist}
\end{equation}
Note that although the above equation uses
the same symbols $T_{\nu_\alpha}$ and $\eta_{\nu_\alpha}$ as Eq.~(\ref{disn}),
the physical meanings of these symbols are very different. In Eq.~(\ref{disn}),
$T_{\nu_\alpha}$ and $\eta_{\nu_\alpha}$ are simple parameters used to fit
the energy distribution functions obtained from supernova neutrino transport
calculations, while in Eq.~(\ref{dist}), $T_{\nu_\alpha}$ is the temperature
and $\eta_{\nu_\alpha}$ is the degeneracy parameter of the $\nu_\alpha$ gas
in the early universe. However, as the assumed
neutrino energy distribution functions for the supernova and
early universe environments have the same functional form,
we use the same symbols in both cases
for convenience. For the homogeneous and isotropic neutrino gas in the early 
universe, the local proper number density of $\nu_\alpha$ is 
\begin{equation}
n_{\nu_\alpha}={{T_{\nu_\alpha}^3}\over{2\pi^2}}F_2(\eta_{\nu_\alpha}).
\label{nodens}
\end{equation}
Equation~(\ref{dist}) can be rewritten as
\begin{equation}
dn_{\nu_\alpha}=n_{\nu_\alpha} {\left( {{d\Omega_\nu}\over{4\pi}} \right)} 
f_{\nu_\alpha}\left(E_\nu\right)dE_\nu.
\label{eudist}
\end{equation}
In terms of the scaled neutrino energy $\epsilon \equiv E_\nu/T_{\nu_\alpha}$,
\begin{equation}
f_{\nu_\alpha}\left(E_\nu\right)dE_\nu=f_{\nu_\alpha}(\epsilon)d\epsilon
={1\over F_2\left(\eta_{\nu_\alpha}\right)}{\epsilon^2d\epsilon\over 
e^{\epsilon-\eta_{\nu_\alpha}}+1}.
\end{equation}
In analogy to the baryon-to-photon ratio 
$\eta\equiv \left(n_b-n_{\bar b}\right)/n_\gamma \approx 6\times {10}^{-10}$,
we define a $\nu_\alpha$-to-photon ratio
\begin{equation}
{\ell}_{\nu_\alpha}\equiv{{n_{\nu_\alpha}-n_{\bar\nu_\alpha}}\over{n_\gamma}} 
={{{\pi^2}\over{12\zeta\left( 3\right)}}}
{\left( {{T_{\nu_\alpha}}\over{T_{\gamma}}} \right)}^3 
\left(\eta_{\nu_\alpha}+{\eta_{\nu_\alpha}^3\over\pi^2}\right),
\label{lepno}
\end{equation}
where $T_\gamma$ is the photon temperature, $\zeta\left(3\right)=1.20206$,
and we have used $T_{\bar\nu_\alpha}=T_{\nu_\alpha}$ and 
$\eta_{\bar\nu_\alpha}=-\eta_{\nu_\alpha}$ to obtain the last identity. 
For $T_{\nu_\alpha}=T_\gamma$ and small $\ell_{\nu_\alpha}$,
$\eta_{\nu_\alpha} \approx 1.46 \ell_{\nu_\alpha}$. Current limits on all 
lepton numbers are $\ell_{\nu_\alpha} < 0.1$ ({\it cf.} Ref. \cite{aba}).

\subsection{Neutrino Propagation in Medium\label{sec-nuprop}}

For a neutrino originating as a $\nu_\alpha$ at $t=0$, its subsequent
flavor evolution along a radially-directed trajectory with Affine parameter 
$t$ is described by
\begin{equation}
i{{\partial}\over{\partial t}}\vert\Psi_{\nu_\alpha}\rangle=
\left({\hat H_{\rm vac}}+{\hat H_{\rm e\nu}}+{\hat H_{\nu\nu}}\right) 
\vert\Psi_{\nu_\alpha}\rangle,
\label{MeanSchroe}
\end{equation}
where we have decomposed the overall evolution Hamiltonian into contributions 
from vacuum neutrino masses and from mean-field ensemble averages 
for neutrino-electron and neutrino-neutrino forward scattering. These
contributions are discussed individually below.

For a neutrino with energy $E_\nu$ and vacuum mass $m\ll E_\nu$, we have 
$E_\nu=\sqrt{p^2+m^2}\approx p+m^2/(2p)$, where $p$ is the magnitude of the
neutrino momentum ${\bf p}$. In this limit, the vacuum-mass contribution to 
the flavor evolution Hamiltonian is
\begin{equation}
{\hat H_{\rm vac}}\approx p{\hat I} + {{1}\over{2p}}
{\left( m_1^2 \vert\nu_1\rangle\langle\nu_1\vert +  
m_2^2 \vert\nu_2\rangle\langle\nu_2\vert\right)},
\label{vacprop}
\end{equation}
where ${\hat I}$ is the identity operator.

Electron neutrinos and antineutrinos can forward scatter on electrons
and positrons through exchange of $W^\pm$. In contrast, there is no such
charged-current forward scattering for $\nu_\mu$, $\nu_\tau$, and their
antiparticles due to the absence of $\mu^\pm$ and $\tau^\pm$ in the 
environments of interest here. Consequently, the effective contribution
from charged-current neutrino-electron forward scattering to the flavor
evolution Hamiltonian is
\begin{equation}
{\hat H_{\rm e\nu}}\left(t\right)=A(t)\vert\nu_e\rangle\langle\nu_e\vert,
\label{ccham}
\end{equation}
where 
\begin{equation}
A(t) \equiv \sqrt{2}G_{\rm F}\left(n_{e^-}-n_{e^+}\right)=\sqrt{2}G_{\rm F} n_b Y_e. 
\end{equation}
In the above equation, $n_{e^-}$, $n_{e^+}$, and $n_b$ are the proper number 
densities of electrons, positrons, and baryons, respectively, at the
position corresponding to time $t$, and 
$Y_e=\left(n_{e^-}-n_{e^+}\right)/n_b$ is the net electron number per 
baryon, or electron fraction.

For a specific neutrino with momentum ${\bf p}$, the effective 
neutral-current neutrino-neutrino forward scattering contribution \cite{FMWS}  
to the flavor evolution Hamiltonian is
\begin{equation}
{\hat H_{\nu\nu}}\left(t\right)=\sqrt{2}G_{F}\int\left(1-\cos\theta_{\bf pq}\right)
\left[{\hat \rho}_{\bf q}(t)-{\hat{\bar\rho}}_{\bf q}(t)\right]d^{3}{\bf q},
\label{nunuham}
\end{equation}
where ${\bf q}$ is the momentum of the background neutrinos and 
$\cos\theta_{\bf pq}= {\bf p}\cdot{\bf q}/pq$.
The term $\left(1-\cos\theta_{\bf pq}\right)$ stems from the structure of 
the weak current \cite{FMWS}. This can be seen from the limit where completely 
relativistic neutrinos are traveling in the same direction along the same 
spacetime path. In this limit $1-\cos\theta_{\bf pq}=0$ and neutrinos
never forward scatter on one another. Obviously, for the homogeneous and 
isotropic neutrino distribution functions characteristic of the early 
universe, $\cos\theta_{\bf pq}$ averages to zero and the 
ensemble average of $\left(1-\cos\theta_{\bf pq}\right)$ is unity.
In the supernova environment the term $\left(1-\cos\theta_{\bf pq}\right)$
will be largest close to the neutron star, where the neutrino trajectories 
can intersect at high angles. At sufficiently large radii above the 
neutron star, the neutrino-neutrino 
forward-scattering contribution to the flavor evolution Hamiltonian
will scale as $r^{-4}$. As the neutrino-electron forward-scattering 
contribution will scale roughly as $r^{-3}$, it may be dominated by
the neutrino-neutrino forward-scattering contribution at small
to moderate distances from the neutron star.

In matrix form, the neutrino flavor evolution equation in the flavor 
basis is
\begin{widetext}
\begin{equation}
i{{\partial\Psi_f}\over{\partial t}}=\left[{\left(p+{{m_1^2+m_2^2}\over{4p}}+{{A}\over{2}}+
\alpha_\nu  \right) {\hat I}+{{1}\over{2}} {\left(\begin{array}{ccc}
A+B-\Delta\cos2\theta & \Delta\sin2\theta+B_{e\tau}\\
 \Delta\sin2\theta+B_{\tau e} & \Delta\cos2\theta-A-B\end{array}\right)}}\right] \Psi_f,
\label{flavorbasis}
\end{equation}
where we have separated the Hamiltonian into a traceless term and a term 
proportional to the identity matrix. The latter term gives only an overall 
phase to the neutrino states, and is therefore unimportant in neutrino 
flavor conversion. In the above equation, $\Delta \equiv \delta m^2/2E_\nu$,
and $\alpha_\nu$, $B$, and $B_{e\tau}$ ($B_{\tau e} = B_{e\tau}^\dagger$) are
the potentials from neutrino-neutrino forward scattering. Specifically,
\begin{eqnarray}
\alpha_\nu & = & {{\sqrt{2}}\over{2}} G_{\rm F} \int\left(1-\cos\theta_{\bf pq}\right)
{\left( {\left[{\hat \rho}_{\bf q}(t)-{\hat{\bar\rho}}_{\bf q}(t)\right]}_{ee}+
{\left[{\hat \rho}_{\bf q}(t)-{\hat{\bar\rho}}_{\bf q}(t)\right]}_{\tau\tau}\right)}d^{3}{\bf q},
\label{alphanu}
\\
B & = & \sqrt{2} G_{\rm F} \int\left(1-\cos\theta_{\bf pq}\right)
{\left( {\left[{\hat \rho}_{\bf q}(t)-{\hat{\bar\rho}}_{\bf q}(t)\right]}_{ee}-
{\left[{\hat \rho}_{\bf q}(t)-{\hat{\bar\rho}}_{\bf q}(t)\right]}_{\tau\tau}\right)}d^{3}{\bf q},
\label{B}
\\
B_{e\tau} & = &2\sqrt{2} G_{\rm F} \int\left(1-\cos\theta_{\bf pq}\right)
{\left[{\hat \rho}_{\bf q}(t)-{\hat{\bar\rho}}_{\bf q}(t)\right]}_{e\tau}d^{3}{\bf q},
\label{Betau}
\end{eqnarray}
where the matrix elements of the density operators are defined as
\begin{eqnarray}
{\left[{\hat\rho}_{\bf q}(t)-{\hat{\bar\rho}}_{\bf q}(t)\right]}_{ee}d^{3}{\bf q}
 & \equiv & 
 \langle \nu_e\vert {\hat \rho}_{\bf q}(t)d^{3}{\bf q}\vert\nu_e\rangle
 -\langle \bar\nu_e\vert {\hat {\bar\rho}}_{\bf q}(t)d^{3}{\bf q}\vert\bar\nu_e\rangle,
 \label{medef1}
 \\
 {\left[{\hat \rho}_{\bf q}(t)-{\hat{\bar\rho}}_{\bf q}(t)\right]}_{\tau\tau}d^{3}{\bf q}
 & \equiv & 
 \langle \nu_\tau\vert {\hat \rho}_{\bf q}(t)d^{3}{\bf q}\vert\nu_\tau\rangle
 -\langle \bar\nu_\tau\vert {\hat {\bar\rho}}_{\bf q}(t)d^{3}{\bf q}\vert\bar\nu_\tau\rangle,
 \label{medef2}
\\
{\left[{\hat \rho}_{\bf q}(t)-{\hat{\bar\rho}}_{\bf q}(t)\right]}_{e\tau}
d^{3}{\bf q}
 & \equiv & 
 \langle \nu_e\vert {\hat \rho}_{\bf q}(t)d^{3}{\bf q}\vert\nu_\tau\rangle
 -\langle \bar\nu_e\vert {\hat {\bar\rho}}_{\bf q}(t)d^{3}{\bf q}\vert\bar\nu_\tau\rangle.
\label{medef}
\end{eqnarray}
\end{widetext}
The physical interpretation of these matrix elements is straightforward, 
even if the notation is cumbersome. For example,
${\left[{\hat \rho}_{\bf q}(t)-{\hat{\bar\rho}}_{\bf q}(t)\right]}_{ee}
d^{3}{\bf q}$ gives the expectation value for the net $\nu_e$ number density
in the pencil of momenta and directions $d^3{\bf q}$ centered on ${\bf q}$.
Note that the off-diagonal matrix element vanishes and makes no contribution 
to $B_{e\tau}$ if neutrinos remain in their initial flavor states.
This is evident if we expand out the first term in Eq.~(\ref{medef}):
\begin{equation}
 \langle \nu_e\vert {\hat \rho}_{\bf q}(t)d^{3}{\bf q}\vert\nu_\tau\rangle =
 \sum_\alpha{ dn_{\nu_\alpha} \langle\nu_e\vert\Psi_{\nu_\alpha}\rangle
 \langle\Psi_{\nu_\alpha}\vert\nu_\tau\rangle}.
\label{expand}
\end{equation}
In the above equation, one or the other amplitude in the sum on the 
right-hand side will be zero unless some neutrino flavor transformation 
has occurred at the time $t$ when this matrix element is evaluated. 
 
For real $B_{e\tau} = B_{\tau e}$, it is convenient to define the effective 
mixing angle $\theta_M$ in medium by
\begin{eqnarray}
\cos2\theta_M\left(t\right) & \equiv &(\Delta\cos2\theta-A-B)/\Delta_{\rm eff},
\label{deleff1}
\\
\sin2\theta_M\left(t\right) & \equiv &(\Delta\sin2\theta+B_{e\tau})/\Delta_{\rm eff},
\label{deleff2}
\end{eqnarray}
where
\begin{equation}
\Delta_{\rm eff}=\sqrt{ {\left( \Delta\cos2\theta-A-B \right)}^2+
{\left( \Delta\sin2\theta+B_{e\tau}\right)}^2 }.
\label{Deltaeff}
\end{equation}
With the term proportional to the identity matrix dropped,
Eq.~(\ref{flavorbasis}) can be transformed to the instantaneous energy basis 
to give
\begin{equation}
i{{\partial\Psi_M}\over{\partial t}}={\left[\begin{array}{cc}
-\Delta_{\rm eff}/2 & -i\dot\theta_M\left(t\right) \\
i\dot\theta_M\left(t\right) & \Delta_{\rm eff}/2\end{array}\right]}\Psi_M,
\label{massbasis}
\end{equation}
where $\dot\theta_M\left(t\right)=d\theta_M/dt$.
In the limit where $|\dot\theta_M\left(t\right)|\ll\Delta_{\rm eff}/2$,
Eq.~(\ref{massbasis}) becomes two decoupled equations and flavor amplitude 
evolution is adiabatic. The flavor evolution equations and the 
corresponding effective mixing angle $\bar\theta_M$ in medium for
antineutrinos can be obtained from those for neutrinos by replacing
$A$, $B$, and $B_{e\tau}$ in the latter with $-A$, $-B$, and $-B_{e\tau}$, 
respectively.
 
The condition $|\dot\theta_M\left(t\right)|\ll\Delta_{\rm eff}/2$ for
adiabatic neutrino flavor evolution is most stringent when
$\Delta_{\rm eff}$ reaches the minimum value $|\Delta\sin2\theta+B_{e\tau}|$
at an MSW resonance corresponding to
\begin{equation}
 \Delta \cos2\theta=A+B.
 \label{res}
 \end{equation}
At resonance, the effective in-medium mixing angle is 
$\theta_M\left( t_{\rm res}\right)=\pi/4$ and mixing is maximal 
with $\sin^22\theta_M=1$. We can define an adiabaticity parameter 
\begin{equation}
\gamma \equiv {{\Delta_{\rm eff}(t_{\rm res})}\over
{2\vert\dot\theta_M\left(t_{\rm res}\right) \vert }}
={(\Delta\sin2\theta+B_{e\tau})^2\over\Delta\cos2\theta}{\cal H},
\label{gamma}
\end{equation}
where
\begin{equation}
{\cal H} \equiv {\left|{{V}\over{\dot{V}}}\right|}_{\rm res}=
\left|{{A+B}\over{\dot{A}+\dot{B}}}\right|_{\rm res}
\label{scale}
\end{equation}
is the scale height for the total potential $V\equiv A+B$ at resonance
with $\dot{V}=dV/dt$. We can gain more insight into 
the adiabaticity parameter by further defining a resonance region
corresponding to $1/2\leq\sin^22\theta_M\leq 1$. In this region the 
change in $V$ around the resonance value $\Delta\cos2\theta$ is
$\delta V=|\Delta\sin2\theta+B_{e\tau}|$, so the width of this region is
\begin{equation}
(\delta t)_{\rm res}={\delta V\over|\dot V|_{\rm res}}=
{|\Delta\sin2\theta+B_{e\tau}|\over\Delta\cos2\theta}{\cal H}.
\end{equation}
As the oscillation length at resonance is
\begin{equation}
L_{\rm res}={{2\pi}\over{\Delta_{\rm eff}}(t_{\rm res})}=
{{2\pi}\over{|\Delta\sin2\theta+B_{e\tau}|}},
\label{osclength}
\end{equation}
we have
\begin{equation}
\gamma=2\pi {(\delta t)_{\rm res}\over{L_{\rm res}}}.
\label{altgamma}
\end{equation}
In summary, large $B_{e\tau}$ increases $\gamma$ in two ways:
(1) by increasing the resonance width $(\delta t)_{\rm res}$ and; (2) by decreasing
the oscillation length $L_{\rm res}$.

Clearly, neutrino flavor evolution will be adiabatic for $\gamma \gg 1$. 
For the small effective vacuum mixing angle $\theta\ll 1$ of interest here, 
neutrino flavor conversion will be complete in this limit. For arbitrary
$\gamma$, the probability of neutrino flavor conversion after propagation 
through resonance is well approximated by $1-P_{\rm LZ}$, where
$P_{\rm LZ}=\exp{\left(-\pi\gamma/2\right)}$ is the Landau-Zener 
probability for a neutrino to jump from one energy eigenstate to the
other in traversing the resonance region. 
 
\subsection{Neutrino Potentials in the Adiabatic Limit}
Due to the cross terms in the single neutrino density operator in
Eq.~(\ref{singleop}), $B_{e\tau}$ is generally
complex. If these cross terms are unimportant, then both $B$ and 
$B_{e\tau}$ are real and their expressions in Eqs.~(\ref{B}) and (\ref{Betau}) 
can be simplified as
\begin{widetext}
\begin{eqnarray}
B&=&-\sqrt{2} G_{\rm F} \sum_{\alpha}{\int{\left(1-\cos\theta_{\bf pq} \right)
{\left[\cos2\theta_M\left(1-2{\vert a_{1\alpha}\vert}^2\right)
dn_{\nu_\alpha}-\cos2\bar\theta_M\left(1-2{\vert\bar{a}_{1\alpha}\vert}^2
\right) dn_{\bar\nu_\alpha}\right]} } },\\
\label{nocross}
B_{e\tau}&=&\sqrt{2} G_{\rm F} \sum_{\alpha}{\int{\left(1-\cos\theta_{\bf pq} \right)
{\left[ \sin2\theta_M\left(1-2{\vert a_{1\alpha}\vert}^2\right) 
dn_{\nu_\alpha}-\sin2\bar\theta_M\left(1-2{\vert \bar{a}_{1\alpha}\vert}^2
\right) dn_{\bar\nu_\alpha} \right]} } }.
\label{nocrossoff}
\end{eqnarray}
As mentioned in Sec.~\ref{sec:nuden}, the cross terms in Eq.~(\ref{singleop})
vanish if neutrino states evolve adiabatically. In this limit, the above
expressions of $B$ and $B_{e\tau}$ can be simplified further. As the 
electron and neutrino number densities at the neutrino sphere are far
above those satisfying the resonance condition, $\nu_e$ and $\nu_\tau$
are born essentially as the energy eigenstates $|\nu_2\rangle$ and 
$|\nu_1\rangle$, respectively. For adiabatic evolution, 
${\vert a_{1e}\vert}^2\approx 0$ and ${\vert a_{1\tau}\vert}^2\approx 1$
for all subsequent time $t$. For antineutrinos, adiabatic evolution gives
${\vert\bar a_{1e}\vert}^2\approx 1$ and 
${\vert\bar a_{1\tau}\vert}^2\approx 0$. Thus, we have
\begin{eqnarray}
B &\approx& \sqrt{2} G_{\rm F} \int{\left(1-\cos\theta_{\bf pq} \right)
{\left[ {\left(dn_{\nu_\tau}-dn_{\nu_e} \right) }\cos2\theta_M + 
{\left(dn_{\bar\nu_\tau}-dn_{\bar\nu_e}\right)}
\cos2\bar\theta_M\right]}},\label{ad2}\\
B_{e\tau} &\approx& \sqrt{2} G_{\rm F} \int{\left(1-\cos\theta_{\bf pq} \right)
{\left[ {\left(dn_{\nu_e}-dn_{\nu_\tau} \right) }\sin2\theta_M + 
{\left(dn_{\bar\nu_e}-dn_{\bar\nu_\tau} \right)}\sin2\bar\theta_M\right]}}.
\label{ad2Betau}
\end{eqnarray}
\end{widetext}

\section{Flavor Mixing with Large Off-Diagonal Potential}

We now focus on adiabatic neutrino flavor evolution, for which the 
potentials from neutrino-neutrino forward scattering are given by
Eqs.~(\ref{ad2}) and (\ref{ad2Betau}). Our main concern is the effects of these
potentials on neutrino flavor evolution in supernovae. 
In Sec.~\ref{sec-nuprop} we have outlined this evolution for
a specific neutrino as it propagates outward from the neutrino sphere.
At a given radius $r$ with potentials $A$ and $B$, the resonance condition 
in Eq.~(\ref{res}) will be met for a particular neutrino energy 
$E_{\rm res}$, i.e.,
\begin{equation}
{\delta m^2\over 2E_{\rm res}}\cos2\theta=A+B.
\label{enures}
\end{equation}
As $B$ and $B_{e\tau}$ at a given radius involve integration of
$\cos2\theta_M$, $\cos2\bar\theta_M$, $\sin2\theta_M$, and
$\sin2\bar\theta_M$ over the neutrino energy distribution functions, it
is convenient to write
\begin{widetext}
\begin{eqnarray}
\cos2\theta_M& = &{1-E_\nu/E_{\rm res}\over{\sqrt{(1-E_\nu/E_{\rm res})^2+
[\tan2\theta+(E_\nu/E_{\rm res})(2E_{\rm res}B_{e\tau})/
(\delta m^2\cos2\theta)]^2}}},
\label{cosM1}
\\
\sin2\theta_M& = &{\tan2\theta+(E_\nu/E_{\rm res})
(2E_{\rm res}B_{e\tau})/(\delta m^2\cos2\theta)
\over{\sqrt{(1-E_\nu/E_{\rm res})^2+
[\tan2\theta+(E_\nu/E_{\rm res})(2E_{\rm res}B_{e\tau})/
(\delta m^2\cos2\theta)]^2}}},
\label{sinM1}
\\
\cos2\bar\theta_M& = &{1+E_\nu/E_{\rm res}\over{\sqrt{(1+E_\nu/E_{\rm res})^2+
[\tan2\theta-(E_\nu/E_{\rm res})(2E_{\rm res}B_{e\tau})/
(\delta m^2\cos2\theta)]^2}}},
\label{cosMbar1}
\\
\sin2\bar\theta_M& = &{\tan2\theta-(E_\nu/E_{\rm res})
(2E_{\rm res}B_{e\tau})/(\delta m^2\cos2\theta)
\over{\sqrt{(1+E_\nu/E_{\rm res})^2+
[\tan2\theta-(E_\nu/E_{\rm res})(2E_{\rm res}B_{e\tau})/
(\delta m^2\cos2\theta)]^2}}}.
\label{sinMbar1}
\end{eqnarray}
\end{widetext}

Let us now consider the limits of the above expressions for the in-medium 
mixing angles when $\theta\ll 1$ and $B_{e\tau}$ is positive and so large 
that the second 
terms in the square root of these expressions dominate the first terms. 
In this limit we have
\begin{eqnarray}
\cos2\theta_M & \to &0,
\label{cosMlim}
\\
\sin2\theta_M & \to &1,
\label{sinMlim}
\\
\cos2\bar\theta_M & \to &0,
\label{cosMbarlim}
\\
\sin2\bar\theta_M & \to &-1,
\label{sinMbarlim}
\end{eqnarray}
for which both neutrinos and antineutrinos have
maximal in-medium mixing with
\begin{eqnarray}
\theta_M & \to & {{\pi}\over{4}},
\label{anglelim1}
\\
\bar\theta_M & \to & {{3\pi}\over{4}}.
\label{anglelim2}
\end{eqnarray}
A large negative $B_{e\tau}$ clearly changes the signs of the limits 
in Eqs.~(\ref{cosMlim})--(\ref{sinMbarlim}), and the
in-medium mixing angles in this case are 
$\theta_M \to {{3\pi}/{4}}$ and $\bar\theta_M  \to  {{\pi}/{4}}$.

In general, we see that large in-medium mixing will
occur simultaneously for neutrinos and antineutrinos over a broad
range of energies if adiabatic flavor evolution results in
\begin{equation}
|B_{e\tau}|\gg{\delta m^2\over 2E_{\rm res}}\cos2\theta
\label{ggbetau}
\end{equation}
at some radius above the neutrino sphere. Using Eq.~(\ref{enures}), we can
rewrite the above equation as $|B_{e\tau}|\gg A+B$. As $\cos2\theta_M\to 0$ 
and $\cos2\bar\theta_M\to 0$ when this is achieved, Eq.~(\ref{ad2}) gives
$B\to 0$. Therefore, Eq.~(\ref{ggbetau}) reduces to
\begin{equation}
|B_{e\tau}|\gg A.
\label{ggbetau2}
\end{equation}

Note that even a $|B_{e\tau}|$ only as large as $(\delta m^2/2E_{\rm res})\cos2\theta$
already has important effects on the in-medium mixing of neutrinos and
antineutrinos. While a neutrino with resonance energy $E_\nu=E_{\rm res}$ 
has maximal in-medium mixing independent of $B_{e\tau}$, the energy range
over which neutrinos have large in-medium mixing with 
$1/2\leq\sin^22\theta_M\leq 1$ is strongly affected by $B_{e\tau}$. For
$B_{e\tau}=0$, this energy range corresponds to
$E_{\rm res}(1-\tan2\theta)\leq E_\nu\leq E_{\rm res}(1+\tan2\theta)$, 
which is very narrow for $\theta\ll 1$. In contrast, for example,
with $B_{e\tau}=(\delta m^2/2E_{\rm res})\cos2\theta$, all neutrinos
with $E_\nu\geq E_{\rm res}/2$ have $1/2\leq\sin^22\theta_M\leq 1$ even 
for $\theta\ll 1$.
Furthermore, $B_{e\tau}$ also affects the in-medium mixing for
antineutrinos, which is strongly suppressed ($\sin^22\bar\theta_M\ll 1$)
in the absence of neutrino-neutrino forward scattering. For
$B_{e\tau}=(\delta m^2/2E_{\rm res})\cos2\theta$ and $\theta\ll 1$,
antineutrinos with $E_\nu\geq E_{\rm res}$ have substantial in-medium
mixing with $1/5\leq\sin^22\bar\theta_M\leq 1/2$. 

\subsection{Towards a Self-Consistent Solution with a Large $B_{e\tau}$}

Here we outline a possible self consistent $B_{e\tau}$-dominant solution 
(BDS) which meets two criteria: (1) $\vert B_{e\tau}\vert \gg A$; and 
(2) adiabaticity, $\gamma \gg 1$. An immediate question is: can adiabatic neutrino flavor evolution 
ever produce a large $B_{e\tau}$ as in Eq.~(\ref{ggbetau2})? Obviously, 
the answer is yes if one can demonstrate that this result is 
obtained under some conditions. Such demonstration requires 
following the flavor evolution of neutrinos with a wide range of
energies covered by their energy distributions. As mentioned in
Sec.~\ref{sec-2}, this process may sound straightforward but turns
out to be computationally difficult. On the other hand, if adiabatic 
flavor evolution can indeed give rise to a large $B_{e\tau}$, then 
the conditions required for this to occur must depend on the following:
the neutrino mixing parameters $\delta m^2$ and $\sin^22\theta$, the
profile of electron number density that gives the potential $A$,
and the neutrino luminosities and energy distribution functions
that are related to the potentials $B$ and $B_{e\tau}$.
Our goal here is to examine these dependences. In so doing, we will
not be able to answer the question posed at the beginning of this
paragraph, but we will be able to provide a range of conditions
that can guide future numerical calculations in search of a
complete solution for neutrino flavor mixing. 

We start with the basic input for our discussion. As explained
in Sec.~\ref{sec-2a}, the mixing parameters of interest here are
$\delta m^2\approx 3\times 10^{-3}$~eV$^2$ and $\sin^22\theta<0.1$.
To characterize the potential $A$, we need the electron number density
$n_e=Y_en_b$. We note that the envelope above 
the post-core-bounce neutron star can be approximated as a quasi-static
configuration with a constant entropy per baryon $S$ in the gravitational 
field of the neutron star. In this case, the enthalpy 
per baryon, $T S$, is roughly the gravitational binding 
energy of a baryon, so that the temperature $T$ scales with radius as
\begin{equation}
T\approx{{M_{\rm NS}\ m_p}\over{m_{\rm Pl}^2}}S^{-1}r^{-1},
\label{enthal}
\end{equation}
where $m_{\rm Pl}\approx 1.221\times{10}^{22}\,{\rm MeV}$ is the Planck 
mass, $m_p$ is the proton mass, and $M_{\rm NS}$ is the neutron star mass.
At late times relevant for $r$-process nucleosynthesis, the environment
above the neutron star is radiation-dominated, so
\begin{equation}
S\approx {2\pi^2\over 45}g_s{T^3\over n_b}
\label{srel}
\end{equation}
in units of Boltzmann constant $k_B$ per baryon. In the above equation,
$g_s$ is the statistical weight in relativistic particles: 
$g_s \approx 11/2$ when $e^\pm$-pairs are abundant and 
$g_s\approx 2$ otherwise. Combining Eqs.~(\ref{enthal}) and (\ref{srel}),
we obtain the run of baryon number density for the $r$-process epoch as
\begin{equation}
n_b\approx {{2\pi^2}\over{45}} g_s
{\left( {{M_{\rm NS} m_p}\over{m_{\rm Pl}^2}}\right)}^3 S^{-4} r^{-3}.
\label{densityrun}
\end{equation}
The potential $A$ is given by
\begin{eqnarray}
A&=&\sqrt{2}G_FY_en_b\approx {{2\sqrt{2}\pi^2}\over{45}} g_sY_eG_F
{\left( {{M_{\rm NS} m_p}\over{m_{\rm Pl}^2}}\right)}^3 S^{-4} r^{-3}
\nonumber\\
&\approx&(5.2\times 10^{-13}\,{\rm MeV})g_sY_e
\left({M_{\rm NS}\over 1.4\,M_\odot}\right)^3S_{100}^{-4}r_6^{-3},
\label{pota}
\end{eqnarray}
where $S_{100}$ is $S$ in units of $100k_B$ per baryon and $r_6$ is $r$
in units of $10^6$~cm.
The $r$-process epoch corresponds to a time post-core-bounce 
$t_{\rm pb}>3\,{\rm s}$. This is a relatively long time 
after core bounce, at least compared with the time scale of the shock 
re-heating epoch and the time scale for evolution of neutrino emission
characteristics such as luminosities and average energies. 
The potential $A$ in Eq.~(\ref{pota}) can also be 
used to describe crudely the {\it shocked} regions of the envelope above 
the core in the shock re-heating epoch, $t_{\rm pb}<1\,{\rm s}$, if we 
take $g_s \sim 1$ and employ a low entropy \cite{QF95}.

To evaluate $B$ and $B_{e\tau}$, we need the differential number density
of each neutrino species at radius $r>R_\nu$ above the neutrino sphere.
The differential $\nu_\alpha$ number density in the absence of flavor 
evolution is given by Eq.~(\ref{normdist}), which depends on the 
luminosity $L_{\nu_\alpha}$ and the average energy 
$\langle E_{\nu_\alpha}\rangle$. For some illustrative numerical estimates we will
assume that all neutrino species have the same luminosity,
\begin{equation}
L_\nu\equiv L_{\nu_e}=L_{\bar\nu_e}=L_{\nu_\tau}=L_{\bar\nu_\tau},
\label{lnu}
\end{equation}
and take the average neutrino energies to be
\begin{eqnarray}
&\langle E_{\nu_e}\rangle=10\ {\rm MeV},\ 
\langle E_{\bar\nu_e}\rangle=15\ {\rm MeV},\nonumber\\ 
&\langle E_{\nu_\tau}\rangle=\langle E_{\bar\nu_\tau}\rangle=27\ {\rm MeV}.
\label{avenu}
\end{eqnarray}

Assuming that adiabatic flavor evolution up to some radius $r>R_\nu$ results
in a BDS described by the criterion in Eq.~(\ref{ggbetau2}), we now examine 
the implications of this criterion for supernova conditions. For definiteness, 
we discuss the case of a large positive $B_{e\tau}^{\rm BDS}$, 
which gives $\sin2\theta_M\to 1$ and
$\sin2\bar\theta_M\to -1$. In this case, Eq.~(\ref{ad2Betau}) gives
\begin{equation}
B_{e\tau}^{\rm BDS}\approx\sqrt{2} G_{\rm F} \int\left(1-\cos\theta_{\bf pq} \right)
(dn_{\nu_e}-dn_{\bar\nu_e}),
\label{ad3Betau}
\end{equation}
where we have assumed that the luminosities and energy distribution functions 
for $\nu_\tau$ and $\bar\nu_\tau$ are very nearly the same in the supernova 
environment. This is a good approximation because these species experience 
nearly identical interactions both in the dense environment of the core 
and in the more tenuous outer regions. For a radially-propagating test
neutrino, the intersecting angles of the background neutrinos, 
$\theta_{\bf pq}$, are coincident with the polar angle in the integration 
over $d\Omega_\nu$ for the test neutrino. Assuming that 
neutrinos of all flavors originate on the same neutrino sphere and using
$dn_{\nu_e}$ and $dn_{\bar\nu_e}$ of the form in Eq.~(\ref{normdist}),
$B_{e\tau}^{\rm BDS}$ in Eq.~(\ref{ad3Betau}) can be evaluated as
\begin{eqnarray}
B_{e\tau}^{\rm BDS}& \approx &  {{\sqrt{2} G_{\rm F}}\over{4\pi R_\nu^2}}
{\left[ 1-\sqrt{1-R_\nu^2/r^2} \right]}^2 
\left( {{L_{\nu_e}}\over{\langle E_{\nu_e}\rangle }}-
{{L_{\bar\nu_e}}\over{\langle E_{\bar\nu_e} \rangle}} \right)\nonumber\\
& \approx & \left(2.1\times{10}^{-10}\,{\rm MeV} \right)R_{\nu 6}^{-2}
{\left(1-\sqrt{1-R_{\nu 6}^2/r_6^2} \right)}^2\nonumber\\
&\times&\left[ {{L_{\nu_e52}}\over{{{\langle E_{\nu_e}\rangle}/{(10\,{\rm MeV})}} }}
-{{L_{\bar\nu_e52}}\over{{{\langle E_{\bar\nu_e}\rangle}/{(10\,{\rm MeV})}}}} 
\right],
\label{SNlimBetau2}
\end{eqnarray}
where $R_{\nu 6}\equiv R_\nu/(10^6\,{\rm cm})$,
$L_{\nu_e52}\equiv L_{\nu_e}/(10^{52}\,{\rm ergs\ s}^{-1})$, and
$L_{\bar\nu_e52}\equiv L_{\bar\nu_e}/(10^{52}\,{\rm ergs\ s}^{-1})$.
For $r\gg R_\nu\sim 10^6\,{\rm cm}$ and the assumptions in Eqs.~(\ref{lnu}) 
and (\ref{avenu}), we obtain
\begin{equation}
B_{e\tau}^{\rm BDS}\approx (1.8\times{10}^{-11}\,{\rm MeV})
{{L_{\nu52}}R_{\nu 6}^2\over{r_6^4}}.
\label{SNlimBetau3}
\end{equation}

Using Eqs.~(\ref{pota}) and (\ref{SNlimBetau2}), we can rewrite the criterion
$B_{e\tau}^{\rm BDS}\gg A$ as
\begin{widetext}
\begin{eqnarray}
{{R_\nu^3/r^3}\over{{\left( 1-\sqrt{ 1-R_\nu^2/r^2 }\right)}^2  }} &\ll& 
{{45}\over{8\pi^3}}{\left( {{ m_{\rm Pl}^2}\over{M_{\rm NS}\, m_p}}\right)}^3 
{\left({{R_\nu S^4}\over{g_s Y_e}}\right)} 
\left( {{L_{\nu_e}}\over{\langle E_{\nu_e}\rangle }}-
{{L_{\bar\nu_e}}\over{\langle E_{\bar\nu_e} \rangle}} \right) 
\nonumber\\
&\approx&{\left( 398\right)} {\left({{1.4\,{\rm M_\odot}}\over{M_{\rm NS}}}\right)}^3
{\left({{R_{\nu 6} S^4_{100}}\over{g_s Y_e}}\right)}
\left[ {{L_{\nu_e52}}\over{{{\langle E_{\nu_e}\rangle}/{(10\,{\rm MeV})}} }}
-{{L_{\bar\nu_e52}}\over{{{\langle E_{\bar\nu_e}\rangle}/{(10\,{\rm MeV})}}}}\right].
\label{rr2}
\end{eqnarray}
\end{widetext}
For $r\gg R_\nu$ and the assumptions in Eqs.~(\ref{lnu}) and (\ref{avenu}), 
we obtain
\begin{equation}
r_6 \ll 33{\left({{1.4\,{\rm M_\odot}}\over{M_{\rm NS}}}\right)}^3
{\left({{ S^4_{100} L_{\nu 52}}R_{\nu 6}^2\over{g_s Y_e}}\right)}.
\label{limul}
\end{equation}

We have assumed adiabatic neutrino flavor evolution in the above discussion.
Though the adiabaticity of the general flavor evolution can be ascertained 
only with a sophisticated numerical treatment, the BDS clearly 
will not be self-consistent if flavor evolution is not adiabatic 
at the radius where $B_{e\tau}^{\rm BDS}\gg A$ is achieved. On the other
hand, if we can show that at this radius the adiabaticity parameter for
the neutrino with energy $E_{\rm res}$ satisfies $\gamma^{\rm BDS}\gg 1$, 
then the BDS is more likely to be obtained. 
With ${\cal{H}}\approx |A/\dot A|\approx r/3$, this criterion
[see Eq.~(\ref{gamma})] can be rewritten as
\begin{eqnarray}
\gamma^{\rm BDS} & \approx & {{B_{e\tau}^2}\over{A}}{\cal{H}}
\nonumber\\
& \approx & {{15}\over{16\sqrt{2}\pi^4}} 
{\left( {{ m_{\rm Pl}^2}\over{M_{\rm NS}\, m_p}}\right)}^3
{\left( {{{1-\sqrt{ 1-R_\nu^2/r^2 }}}\over{R_\nu/r}} \right)}^4
\nonumber\\
&\times&\left({{G_{\rm F}S^4}\over{g_s Y_e}}\right)
{\left( {{L_{\nu_e}}\over{\langle E_{\nu_e}\rangle }}-
{{L_{\bar\nu_e}}\over{\langle E_{\bar\nu_e} \rangle}} \right)}^2\gg 1,
\label{completead}
\end{eqnarray}
which reduces to
\begin{equation}
\gamma^{\rm BDS}\approx 10^7{\left({{1.4\,{\rm M_\odot}}\over{M_{\rm NS}}}\right)}^3
\left({{S_{100}^4L_{\nu 52}^2}\over{g_s Y_e}}\right){{R^4_{\nu 6}}\over{r_6^4}}\gg 1
\label{appad}
\end{equation}
for $r\gg R_\nu$ and the assumptions in Eqs.~(\ref{lnu}) and (\ref{avenu}).

The criteria in Eqs.~(\ref{rr2}) and (\ref{completead}) 
can be met in some regions with significant 
scale and duration above the neutron star during both the $r$-process and shock 
re-heating epochs. For example, an $r$-process environment with modest entropy 
might have $Y_e\approx 0.4$, $R_{\nu 6}\approx 1$, $g_s\approx 11/2$, 
$S_{100}\approx 1.5$, and $L_{\nu {52}} \approx 0.1$. For these parameters 
Eqs.~(\ref{rr2}) and (\ref{completead}) would give $r_6 \ll 8$ and 22, 
respectively, so the BDS may be obtained over an extended region above the
neutrino sphere. For a higher entropy, $S_{100} = 2.5$, but with all the other
parameters remaining the same, Eqs.~(\ref{rr2}) and (\ref{completead}) would 
give $r_6 \ll 60$ and $40$, respectively.
We can put these limits in perspective by noting the temperature at which
salient events or processes occur above the neutron star. The radius 
corresponding to a temperature $T_9$ (measured in units of ${10}^9\,{\rm K}$) 
is very roughly $r_6 \approx 22.5/(T_9 S_{100})$ [see Eq.~(\ref{enthal})]. 
Weak freeze-out, where the neutron-to-proton ratio is set, occurs 
at $T_9 \sim 10$. The neutron capture regime in the $r$-process is 
typically further out, occuring between $T_9 \approx 3$ and $T_9\approx 1$. 
Therefore, the limits on the radius discussed above are 
so generous that maximal in-medium mixing for both neutrinos and antineutrinos
associated with the BDS 
could affect important weak interaction processes in the envelope, the 
$r$-process, and the neutrino signal.

\begin{figure}
\includegraphics[width=3.25in]{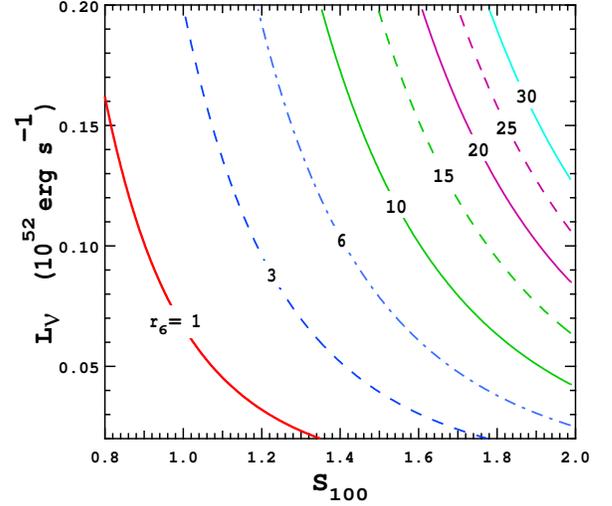}
\caption{ Contours of limiting radius (in units of ${10}^6\,{\rm cm}$)
beneath which $B_{e\tau}^{\rm BDS}\gg A$ may be obtained.
Except for the region corresponding to the larger values of the limiting 
radius, these results are generally more stringent than those from
$\gamma^{\rm BDS}\gg 1$. The chosen range of parameters is
meant to be characteristic of the $r$-process epoch.
The horizontal axis is entropy
in units of $100k_B$ per baryon, while
the vertical axis is neutrino luminosity $L_{\nu}$ in units
of ${10}^{52}\,{\rm ergs}\,{\rm s}^{-1}$. 
Here we take $Y_e =0.4$, $g_s=11/2$, $M_{\rm NS}=1.4\,{\rm M}_\odot$,
and $R_\nu =10\,{\rm km}$. We also assume that 
$L_{\nu_e}=L_{\bar\nu_e}=L_{\nu}$, $\langle E_{\nu_e}\rangle=10\,{\rm MeV}$,
$\langle E_{\bar\nu_e}\rangle=15\,{\rm MeV}$, and that all mu and tau neutrinos
and antineutrinos have identical luminosities and energy spectra.}
\label{figure1}
\end{figure}

\begin{figure}
\includegraphics[width=3.25in]{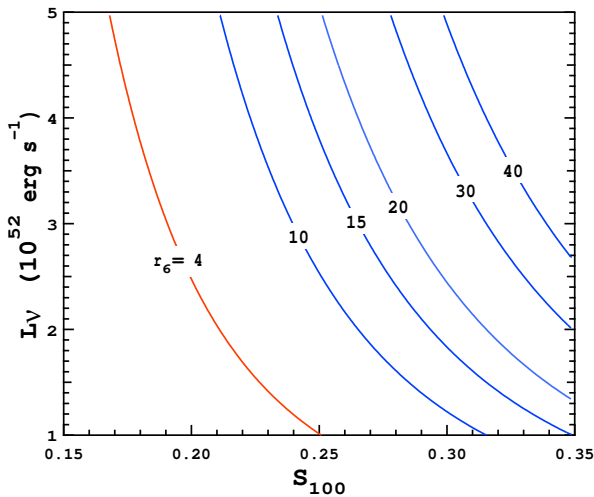}
\caption{Same as Fig.~\ref{figure1}, except that 
now the conditions are meant to be representative
of the shock re-heating epoch. In this case we take
$Y_e =0.35$, $g_s=1.5$, and $R_\nu =40\,{\rm km}$.}
\label{figure2}
\end{figure}

Taking $Y_e =0.4$, $g_s=11/2$, $M_{\rm NS}=1.4\,{\rm M}_\odot$,
$R_\nu =10\,{\rm km}$,
$L_{\nu_e}=L_{\bar\nu_e}=L_{\nu}$, $\langle E_{\nu_e}\rangle=10\,{\rm MeV}$,
and $\langle E_{\bar\nu_e}\rangle=15\,{\rm MeV}$, we use Eq.~(\ref{rr2})
to calculate the combinations of $L_\nu$ and $S$ for which the criterion
$B_{e\tau}^{\rm BDS}\gg A$ can be met below a fixed radius. The results
are shown in Fig.~\ref{figure1} as contours labelled by the limiting radius.
Except for the region corresponding to the larger values of the limiting 
radius, these results are generally more stringent than those from the
criterion $\gamma^{\rm BDS}\gg 1$. Compared with the $r$-process regime at
later times, the shock re-heating epoch is characterized by much higher 
neutrino luminosites. In general both $L_{\nu_e}$ and $L_{\bar\nu_e}$ are 
$\sim {10}^{52}\,{\rm ergs}\,{\rm s}^{-1}$. Taking $Y_e =0.35$, $g_s=1.5$, 
and $R_\nu =40\,{\rm km}$ (other parameters remaining the same as for
Fig.~\ref{figure1}), we present in Fig.~\ref{figure2} the constraints on 
$L_{\nu}$ and $S$ for which $B_{e\tau}^{\rm BDS}\gg A$ can be met below 
various radii during the shock re-heating epoch. Based on these results,
both shock re-heating and the neutrino signal could be 
affected by maximal neutrino flavor mixing \cite{FM,SF}
if there were a hierarchy of neutrino energies at this epoch. 

Note that average $\nu_e$ and $\bar\nu_e$ energies may be quite similar 
during much of the shock re-heating epoch, but the luminosities for 
$\nu_e$ can be significantly larger than those for $\bar\nu_e$. This is 
especially true for shock break-out through the neutrino sphere, the 
so-called neutronization burst. For a time span of $\sim 10$ ms 
we could have $L_{\nu_e 52} \sim 10$, while $L_{\bar\nu_e 52}$ 
is an order of magnitude smaller. Since neutrino flavor mixing in 
the coherent limit is a phase effect, the $10\,{\rm ms}$ duration 
of this high-luminosity burst may be long enough to establish the 
BDS. Neglecting $L_{\bar\nu_e}$ and taking $L_{\nu_e 52}=10$, 
$\langle E_{\nu_e}\rangle =10\,{\rm MeV}$, $R_{\nu 6}=4$, $Y_e=0.35$, 
$g_s=1.5$, and $S_{100}=0.15$, we find that 
$B_{e\tau}^{\rm BDS}\gg A$ may be obtained for $r_6\ll 15$
[Eq.~(\ref{rr2})]. This limit becomes $r_6\ll 50$ if $S_{100}=0.2$. 
For both cases the limit from Eq.~(\ref{completead}) is much weaker.
Therefore, the neutrino signal from the neutronization 
burst and the early shock re-heating process could be 
affected by maximal neutrino flavor mixing associated with the BDS. 

\subsection{Is the BDS Ever Attained?}

Achieving the BDS is dependent on a number of conditions, many of which are unlikely to strictly and generally obtain in environments in nature with high neutrino fluxes. The essence of the 
BDS is the dominance of the flavor off-diagonal potential and, in particular, $2E_\nu B_{e\tau} \gg \delta m^2 \cos2\theta$. Since the measured neutrino mass-squared differences are small, it will not take a large flavor off-diagonal potential to force the system into something like the 
BDS. 

However, as Eq.\ (\ref{expand}) shows, a necessary condition for $B_{e\tau}$ to be non-zero at some time/position is that {\it some} neutrinos must have transformed their flavors there. This can be problematic because in both the early universe and the post-shock supernova environment a fluid element will evolve from conditions of very high density toward lower density. For example, the region near the neutron star surface is very high density, corresponding to high electron degeneracy. This will tend to suppress in-medium neutrino mixing. A hydrodynamic flow away from the neutron star surface will carry a fluid element into regions of lower temperature and density and net neutrino fluxes. At large enough radius the neutrino-electron potential will scale like $A\sim r^{-3}$, while the flavor-diagonal and off-diagonal neutrino-neutrino potentials will scale as $r^{-4}$. As a result, there may be some region where the neutrino-neutrino potentials dominate.

The neutrino resonance energy experienced in this fluid element at radius $r$ will be  $E_{\rm res} = \delta m^2\cos2\theta/2(A+B)$. Near the neutron star surface $E_{\rm res}$ will be extremely small. Further out, in an adiabatic and roughly hydrostatic envelope (notation as in the last section), the resonance energy at radius $r$ will be
\begin{widetext}
\begin{eqnarray}
E_{\rm res} & \approx & {{45}\over{4\sqrt{2} \pi^2}} {\left( {{ m_{\rm Pl}^2}\over{M_{\rm NS}\, m_p}}\right)}^3
{{\delta m^2\cos2\theta\ S^4\ r^3}\over{G_{\rm F} \left(Y_e+Y_\nu^{\rm eff}\right)}}
\label{eres1}
\\
& \approx & \left( 2.85\times{10}^{-3}\,{\rm MeV} \right) {\left( {{1.4\,{\rm M}_\odot}\over{M_{\rm NS}}} \right)}^3
{\left( {{\delta m^2\cos2\theta}\over{3\times{10}^{-3}\,{\rm eV}^2}} \right)} {{S_{100}^4\ r_6^3}\over{ \left(Y_e+Y_\nu^{\rm eff}\right) }},
\label{eres2}
\end{eqnarray}
\end{widetext}
where we define the effective net number of neutrinos per baryon through $B=\sqrt{2} G_{\rm F} n_{\rm b} Y_\nu^{\rm eff}$. Because $\delta m^2$ is small, the resonance energy also tends to be small at distances where neutrino fluxes are appreciable.

\begin{figure}
\includegraphics[width=3.25in]{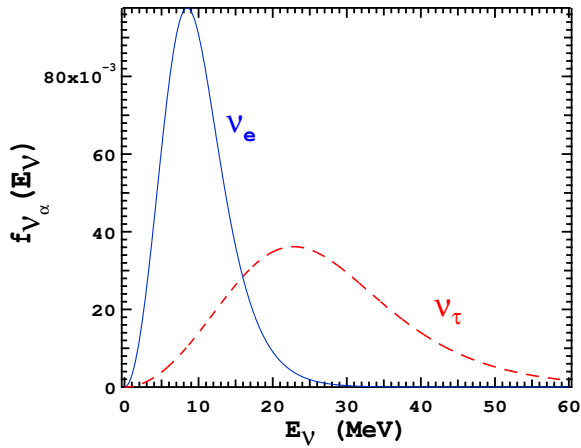}
\caption{Example of normalized energy distribution functions for $\alpha = e,\ \tau$ neutrinos at the neutrino sphere in the supernova environment. Here we take $\langle E_{\nu_e}\rangle=10\,{\rm MeV}$ and $\langle E_{\nu_\tau}\rangle=27\,{\rm MeV}$ and the neutrino degeneracy parameter for both flavors to be $\eta_{\nu_\alpha} =3$.}
\label{figure3}
\end{figure}

However, if neutrinos transform their flavors via a strict MSW evolution then $B$ (and $Y_\nu^{\rm eff}$) will drop with the radius of the fluid element and eventually will be driven {\it negative} \cite{QF95}. To see this consider first an example (hierarchical) energy spectrum for $\nu_e$ and $\nu_\tau$ neutrinos as they leave the neutrino sphere. In Fig.\ (\ref{figure3}) we show Fermi-Dirac-type energy spectra for these species, taking the neutrino degeneracy parameter for both to be $\eta_{\nu_\alpha} =3$, and taking average energies $\langle E_{\nu_e}\rangle = 10\,{\rm MeV}$ and $\langle E_{\nu_\tau}\rangle = 27\,{\rm MeV}$. The actual supernova neutrino energy spectra may differ significantly from these, but they serve to illustrate general trends. 
Note that for our chosen spectral parameters, the $\nu_e$ population at lower energies is {\it larger} than the $\nu_\tau$ population for comparable luminosity in the two neutrino species.

\begin{figure}
\includegraphics[width=3.25in]{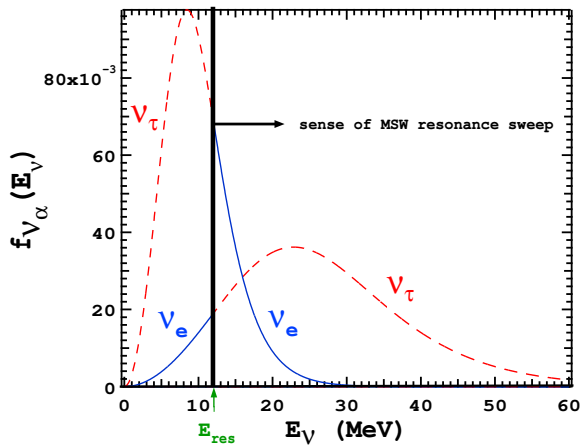}
\caption{Neutrino energy distribution functions in a fluid element at some time $t$ corresponding to position $r$. These have the same neutrino energy/temperature parameters as in the previous figure. Here, however, complete flavor conversion between $\nu_e$ and $\nu_\tau$ has taken place from $E_\nu=0$ to the MSW resonance energy at this time/position $E_{\rm res} = 12\,{\rm MeV}$.
As the fluid element moves out the resonance energy will increase and sweep from left to right through the neutrino distributions}
\label{figure4}
\end{figure}

As our example fluid element moves out to larger $r$, the resonance energy will also increase. It could increase significantly if $\vert A+B\vert \rightarrow 0$. If neutrino flavor conversion in the channel $\nu_e\rightleftharpoons \nu_\tau$ is efficient and complete, then at some point we will have the situation depicted in Fig.\ (\ref{figure4}). Here $B$ could be negative because we have swapped flavors at low neutrino energy and, for our chosen spectral parameters, the $\nu_e$ population now may be {\it smaller} than the $\nu_\tau$ population. Furthermore, in this situation the material may be driven more neutron-rich (lower $Y_e$) on account of the now altered competition between the processes in Eqs.\ (\ref{rates1}) \& (\ref{rates2}). Eventually, of course, the resonance will sweep through the higher energy regions of the distribution functions and the fluid element will move further out to where neutrino fluxes are lower.

The higher the resonance energy, the greater the neutrino population which has been appreciably mixed and, hence, the larger will be $B_{e\tau}$. The importance of this can be ascertained by comparing $B_{e\tau}^{\rm BDS}$ to the vacuum term
\begin{widetext}
\begin{equation}
{{\delta m^2\cos2\theta}\over{2E_\nu}} \approx \left(1.5\times{10}^{-16}\,{\rm MeV} \right) {\left({{\delta m^2\cos2\theta}\over{3\times{10}^{-3}\,{\rm eV}^2}}  \right)}  {\left( {{10\,{\rm MeV}}\over{E_\nu}}  \right)}.
\label{vacpot}
\end{equation}
\end{widetext}
From Eq.\ (\ref{SNlimBetau2}) it is clear that this term could be substantially smaller than $B_{e\tau}^{\rm BDS}$ if $E_\nu$ is a typical neutrino energy. Even if this is not true for $E_\nu=E_{\rm res}$ at very high density where $E_{\rm res}$ is small, higher energy neutrinos and antineutrinos may experience significant in-medium mixing angles over a broad range of energy. Though not strictly our 
BDS, this may nevertheless approximate it.

Previous numerical simulation work on neutrino flavor evolution in the supernova environment may offer only limited guidance here. The simulation in Ref. \cite{QF95} made the same $2\times 2$, and one-dimensional approximations as we make here. (By \lq\lq one-dimensional\rq\rq\ we mean that flavor histories on neutrino trajectories of any polar angle are taken to be the same as a radially directed path for the same lapse of Affine parameter along these trajectories.) Additionally, the work in Ref. \cite{QF95} employed the density profiles and neutrino fluxes of the Mayle \& Wilson late time supernova models and it adopted a range for $\delta m^2$ which is now known to be un-physically large for active-active neutrino evolution. Both of these features combined to produce only minimal effects from rather small values of $B_{e\tau}$.

Likewise, the numerical simulation of Ref. \cite{Pastor} considered one-dimensional, $2\times 2$ neutrino flavor evolution with un-physically large mass-squared difference. The conclusions in this work regarding real supernovae are suspect because: (1) the large $\delta m^2$ used would in reality demand the incorporation of sterile neutrinos which mix significantly with actives and this was left out; and (2) the feedback of neutrino flavor conversion on $Y_e$ was not correctly modeled since the threshold was neglected in the rate for $\bar\nu_e+p\rightarrow n+e^+$ \cite{Qian} and the weak magnetism corrections \cite{Horow} were also neglected. However, this numerical simulation was the first to follow neutrino phases in detail in this environment. Synchronization of large amplitude neutrino flavor oscillations was seen. This behavior is at least qualitatively like some aspects of 
the BDS, especially as regards significant in-medium mixing. 

Though the conditions for establishment of the BDS are manifest in many regions of the post-shock supernova environment, it has not been seen unambiguously in simulations to date. However, there is considerable room for improvement in the sophistication of these simulations. Flavor evolution histories on different neutrino trajectories needs to be followed in detail, including all coupling. The role of density fluctuations \cite{density} in getting some neutrino conversion going at high density also needs to be investigated. Likewise, legitimate three-neutrino mixing of neutrinos and antineutrinos must be followed.
Finally, the effects of neutrino mixing on neutrino transport in the neutron star core may be important and recent formulations \cite{Sawyer,Burrows} of this problem represent significant progress.

\subsection{The Ephemeral Nature of the BDS}

Changing neutrino luminosities and fluxes and changing matter density will quickly lead to the development of complex amplitudes in the unitary transformation between the neutrino mass/energy and flavor bases which, in turn, will lead to complex potentials. This will signal the end of the strict validity of our particular 
BDS discussed above. However, it may not signal the immediate end of appreciable in-medium mixing among the flavors of neutrinos and antineutrinos. 

If we ride along with a fluid element being driven from the neutron star's surface by heating we will see a local fall off in matter density and neutrino fluxes and so a decrease in neutrino-electron and neutrino-neutrino forward scattering-induced potentials in this Lagrangian frame. What is the effect of this time dependence on in-medium flavor mixing?
Using the flavor basis evolution equation (Eq.\ \ref{flavorbasis}) and ignoring the term proportional to the identity we can find a second order equation for, {\it e.g.}, $a_{e\alpha}$, the amplitude for a neutrino of initial flavor $\alpha$ to be a $\nu_e$:
\begin{equation}
{\ddot{a}}_{e\alpha} + \omega^2\ a_{e\alpha}= {{\dot{B}_{e\tau}}\over{B_{e\tau}}}\ {\dot{a}}_{e\alpha}.
\label{deflavor}
\end{equation}
Here the dots over quantities denote time derivatives and 
\begin{equation}
\omega^2 = {{1}\over{4}} \left[{\vert B_{e\tau}\vert}^2 +\delta^2 +2i {\dot{\delta}}-2i{{\delta {\dot{B}}_{e\tau}}\over{B_{e\tau}}}  \right],
\label{omegade}
\end{equation}
with $\delta\equiv A+B-\Delta\cos2\theta$ and ${\dot{\delta}}=\dot{A}+\dot{B}$.

In our BDS all time derivatives vanish, $\omega \approx \vert B_{e\tau}\vert/2$,
and we can solve Eq.\ (\ref{deflavor}), use unitarity ($\vert a_{e\alpha}\vert^2+\vert a_{\tau\alpha}\vert^2=1$) and take $a_{ee}=a_{e\tau}=a_{\tau e}=\pm \exp{\left( \pm i \omega t\right)}/\sqrt{2}$ and $a_{\tau \tau}=\mp \exp{\left( \pm i \omega t\right)}/\sqrt{2}$, and likewise, ${\bar{a}}_{ee}={\bar{a}}_{e\tau}={\bar{a}}_{\tau e}=\pm \exp{\left( \pm i \omega t\right)}/\sqrt{2}$ and ${\bar{a}}_{\tau \tau}=\mp \exp{\left( \pm i \omega t\right)}/\sqrt{2}$. If we employ these solutions in the general flavor-basis form for the off-diagonal potential,
\begin{widetext}
\begin{equation}
B_{e\tau}= \sqrt{2}G_{\rm F}\sum_\alpha \int{\left( 1-\cos\theta_{\bf{p q}}\right)\left[ dn_{\nu_\alpha} a_{e\alpha} a^\ast_{\tau\alpha}-dn_{\bar\nu_\alpha} {\bar{a}}_{e\alpha} {\bar{a}}^\ast_{\tau\alpha}\right]},
\label{flavorBetau}
\end{equation}
we will recover the BDS form for this [{\it cf.}, Eq.~(\ref{ad3Betau})] discussed above:
\begin{equation}
B^{\rm BDS}_{e\tau} \approx \sqrt{2}G_{\rm F}\int{\left( 1-\cos\theta_{\bf{p q}}\right)\left[ \left(dn_{\nu_e}-dn_{\bar\nu_e}\right)-\left(dn_{\nu_\tau}-dn_{\bar\nu_\tau}\right)\right]}.
\label{fpflavorlim}
\end{equation}
\end{widetext}
However, once we allow the potentials to change in time, amplitudes will quickly acquire a non-sinusoidal time dependence which will lead to the development of potentials with imaginary components. With complex potentials we will lose a key assumption used in obtaining the 
BDS of Eq.\ (\ref{fpflavorlim}). Flavor evolution from that point on will be complicated, but there is nothing in the evolution equations that demands an immediate return to medium-suppressed flavor mixing for most neutrino energies.

\section{The BDS in Lepton-Degenerate Cosmologies}

Coherent active-active neutrino flavor evolution in the early universe also can be dominated by the flavor off-diagonal potential whenever significant net lepton numbers reside in the neutrino seas. Collision-associated decoherence dominates neutrino flavor conversion in the early universe at temperatures above Weak Decoupling, $T> 1\,{\rm MeV}$. Neutrino inelastic scattering rates are large compared to the expansion rate in that regime. By contrast, neutrino flavor evolution below this scale is largely coherent. We will concentrate on this epoch. For illustrative purposes we also will confine our discussion to cosmologies which have identical lepton numbers and interactions for both mu and tau neutrinos. With this condition we can reduce the flavor evolution problem to the same $2\times 2$ channel we dealt with for the supernova environment. 

Coherent medium-enhanced flavor transformation in the channel $\nu_{\alpha}\rightleftharpoons\nu_{\beta}$ (where $\alpha,\beta=e,\mu,\tau$ and $\alpha\neq \beta$) in the early universe is governed by the flavor-basis Hamiltonian
\begin{equation}
{\hat H}_{\rm EU} = {{1}\over{2}} {\left(\begin{array}{ccc}
V-\Delta\cos2\theta & \Delta\sin2\theta+B_{e\tau}\\
 \Delta\sin2\theta+B_{\tau e} & \Delta\cos2\theta-V\end{array}\right)},
\label{euham}
\end{equation}
where $V=A+B+\delta C$, we take $\alpha=e$ and $\beta=\tau$, and we ignore the components of the coherent Hamiltonian proportional to the identity. The time evolution of the flavor and mass/energy amplitudes can be handled by a mean field Schr\"odinger-like equation in complete analogy to Eq.s\ (\ref{flavorbasis}) \& (\ref{massbasis}). 

The thermal contribution to the flavor-diagonal potential is $\delta C$. The high entropy of the universe ($S\sim {10}^{10}$) dictates a sometimes appreciable contribution to neutrino effective mass stemming from forward scattering on thermal fluctuations ({\it cf.} Ref.\ \cite{AFP}).  
The thermal term relative to zero potential is $C\approx r_\alpha G_{\rm F}^2 T^5$, where $r_e \approx 79.34$ and $r_\tau\approx 22.22$ at the epoch of interest, $T<2\,{\rm MeV}$. The difference of these contributions is $\delta C\approx \delta r G_{\rm F}^2 T^5$, where $\delta r =r_e-r_\tau$. This term in the Hamiltonian is negligible in the post Weak Decoupling epoch environment whenever the lepton numbers are significantly larger than the baryon-to-photon ratio, $\vert \ell_{\nu_\alpha}\vert\gg\eta$. We therefore will neglect $\delta C$ here, as we consider the large lepton number case only. Similarly, we also neglect thermal contributions to the flavor off-diagonal potential.

Let us consider the case where $\ell_{\nu_\tau} > \ell_{\nu_e}$. We impose the BDS for this case by: (1) assuming completely adiabatic neutrino flavor evolution; and (2) employing the maximal mixing angles of Eq.s\ (\ref{cosMlim}), (\ref{sinMlim}), (\ref{cosMbarlim}), (\ref{sinMbarlim}). The first of these conditions allows the use of the adiabatic forms for $B$ and $B_{e\tau}$, Eq.s\ (\ref{nocross}) and (\ref{nocrossoff}), respectively. Employing the second assumption in these potential forms immediately implies that $B\approx 0$.

We will have $\vert a_{1\tau}\vert^2=0$, $\vert a_{1e}\vert^2=1$, $\vert\bar a_{1\tau}\vert^2=1$, and $\vert\bar a_{1e}\vert^2=0$ in the adiabatic limit when $\ell_{\nu_\tau} > \ell_{\nu_e}>0$. (There is a level crossing for the antineutrinos in this case but not for the neutrinos.) Employing these amplitudes in Eq.\ (\ref{nocrossoff}) and noting the isotropic nature of the neutrino distribution functions in the early universe, we find
\begin{eqnarray}
B_{e\tau}  & \approx & \sqrt{2} G_{\rm F} {\left[ \left( n_{\nu_\tau}-n_{\bar\nu_\tau}\right)-\left(n_{\nu_e}-n_{\bar\nu_e} \right)  \right]}  
\label{Betaueu}
\\
& \approx & {{2\sqrt{2} \zeta\left( 3\right)}\over{\pi^2}} G_{\rm F} T^3 \left( \ell_{\nu_\tau}-\ell_{\nu_e}\right).
\label{BetaueuL}
\end{eqnarray}

For this solution to be self consistent, we must have $\vert B_{e\tau}\vert \gg A$ and adiabatic neutrino flavor evolution. The first condition will be true so long as
\begin{equation}
\left( \ell_{\nu_\tau}-\ell_{\nu_e}\right) \gg Y_e\ \eta \sim 3\times{10}^{-10}.
\label{eucond}
\end{equation}
This condition follows on noting that $A=\sqrt{2} G_{\rm F} Y_e n_{\rm b}$ and the baryon density is $n_{\rm b} = \eta\ n_{\gamma}$. 

Adiabatic flavor evolution in this case is all but guaranteed on account of the slow expansion rate at this epoch and the known values for the atmospheric and solar mass-squared difference scales (see the discussion of adiabaticity in Ref. \cite{ABFW}). The effective density scale height for weak charge at an epoch with temperature $T$ in the early universe is roughly a third of the causal horizon length or, more correctly, 
\begin{equation}
{\cal{H}} \equiv {\Bigg\vert  {{1}\over{V}} {{dV}\over{dt}}\Bigg\vert}^{-1}\approx {{1}\over{3}} H^{-1}{\Bigg\vert 1+{{\dot{g}/g}\over{3H}}-{{{{\Delta \dot\ell}}/{\Delta \ell}}\over{3H}} \Bigg\vert}^{-1}.
\label{denscaleeu}
\end{equation}
The statistical weight in relativistic particles at this epoch is $g$ and the difference in lepton numbers here is $\Delta \ell\equiv \ell_{\nu_\tau}-\ell_{\nu_e}$. The expansion rate (Hubble parameter) is $H\approx \left( 8\pi^3/90\right)^{1/2} g^{1/2} T^2/m_{\rm Pl}$ and the horizon length in $d_H(t)=2t=H^{-1}$. Neglecting the rate of change of $g$ and the rate of change of lepton difference, ${{\Delta \dot\ell}}$, relative to the expansion rate, the adiabaticity parameter for the 
BDS [cf. Eq.\ (\ref{completead})] is in this case
\begin{eqnarray}
\gamma & \approx & {{\sqrt{10}\ \zeta\left( 3\right) G_{\rm F}\, m_{\rm Pl}}\over{\pi^{7/2}}} \cdot {{T \left(\Delta \ell\right)^2}\over{g^{1/2}\, Y_e\ \eta}} 
\label{gameu1}
\\
& \approx & \left( 5\times{10}^{18}\right) {\left( {{10.75}\over{g}} \right)}^{1/2} {{\left(\Delta \ell\right)^2}\over{Y_e}} {\left( {{T}\over{\rm MeV}} \right)}. 
\label{approxgameu}
\end{eqnarray}
In the second approximation we have used the measured baryon-to-photon ratio $\eta \approx 6\times{10}^{-10}$. For an epoch where $T\sim 1\,{\rm MeV}$, we will have $g\approx 10.75$ and $Y_e\approx 0.5$ to $0.16$, so that we have adiabatic neutrino flavor evolution for $\Delta \ell \gg \eta$. Since this is the same condition as that required for  $\vert B_{e\tau}\vert \gg A$ to obtain, we see that $B_{e\tau}^{\rm BDS}$ indeed can be a self consistent solution in the early universe. 

If we had a lepton number degenerate cosmology with $\ell_{\nu_\tau} >\ell_{\nu_e}$ as described above, then we would expect a rapid \lq\lq evening up\rq\rq\ of the electron and tau lepton numbers if the maximal 
mixing BDS obtained. In fact, numerical simulations which include the neutrino background terms and follow neutrino amplitudes {\it with} phases find just this end result \cite{abb}. They also find that neutrino oscillations are synchronized in phase in time/space. To what extent is this result related to the 
BDS?

In contrast to the supernova case, it may be much easier to evolve into the BDS 
in a large lepton number early universe scenario. First, with modest lepton numbers, $\vert \ell_{\nu_\alpha}\vert < {10}^{-4}$, there could be significant transformation of neutrino flavors associated with collisions and de-coherence in the epoch where temperatures are above the weak decoupling scale. The extent of this conversion depends on vacuum mixing angle $\theta$ and, hence, possibly on $\theta_{13}$. Significant flavor conversion at this epoch would imply that there may already be a significant $\vert B_{e\tau}\vert$ by the time neutrino flavor evolution is dominated by coherent forward scattering in the epoch after weak decoupling. 

Also, depending on $\Delta \ell$, neutrino resonance energies could be appreciable at this epoch. The resonance condition implies that the resonant scaled energy is
\begin{equation}
\epsilon_{\rm res} \approx {{\pi^2\ \delta m^2\cos2\theta}\over{4\sqrt{2}\zeta\left( 3\right) G_{\rm F}}}
\cdot{{1}\over{\left( \Delta \ell + \eta Y_e \right)\ T^4 }}.
\label{euresonance}
\end{equation}
It would not take much coherent flavor transformation to produce a prodigious $B_{e\tau}$, at least on the scale of the vacuum term in Eq.\ (\ref{vacpot}). This term can be especially small here on account of the large resonance energies. For example, using Eq.\ (\ref{euresonance}), the resonance energy $= T \epsilon_{\rm res}$ is
\begin{widetext}
\begin{equation}
E_{\rm res} \approx \left( 6.2\times{10}^{5}\,{\rm MeV}\right) \left({{\delta m^2\cos2\theta}\over{3\times{10}^{-3}\,{\rm eV}^2}}\right) \left({{\rm MeV}\over{T}}\right)^3 {{1}\over{Y_e+\Delta \ell/\eta}}.
\label{resnumbers}
\end{equation}
\end{widetext}
In the BDS only the $A$ term determines resonance energies. This case would correspond to $\Delta \ell=0$ in Eq.\ (\ref{euresonance}). This would give a large resonance energy and would likely lead to a significant $\vert B_{e\tau}\vert$ and appreciable mixing across a broad range of neutrino energies.

As outlined above, the numerical simulations of Ref.s \cite{abb} show 
synchronization of large amplitude neutrino and/or antineutrino flavor 
oscillations in large lepton number early universe scenarios. As for 
the supernova case, this looks at least qualitatively similar to the 
BDS. This is especially true while $\Delta \ell$ is 
still large in these models. As this quantity is decreased by neutrino 
flavor mixing, the flavor off-diagonal potential is decreased and neutrino 
flavor inter-conversion will be less efficient and the BDS 
conditions may no longer apply. 
   
\section{Conclusion}

We have investigated the complicated problem of $2\times 2$ coherent 
neutrino flavor evolution in the limit of large flavor off-diagonal 
neutrino-neutrino forward scattering potential both in the early 
universe and in the post-shock supernova environment. We have 
identified a simple solution/limit in this problem. 
This solution (BDS) is governed by a dominant off-diagonal potential. 
This constitutes a viable solution only under a number of restrictive 
assumptions, but it is evident that even a rough facsimile to this 
solution will retain key principal features of the BDS. These 
include maximal or near maximal in-medium mixing angles for both 
neutrinos and antineutrinos over broad ranges of neutrino/antineutrino 
energies. These features are very different from conventional neutrino 
flavor amplitude evolution with the MSW effect. 

Indeed, it has been generally thought that the small values of neutrino 
mass-squared difference among the active neutrinos preclude significant 
effects from medium-enhanced neutrino flavor conversion in the dense 
environment above the neutron star in supernova models. This {\it is} 
largely the case for conventional MSW neutrino evolution. It need not 
be the case when neutrino-neutrino forward scattering potentials are large. Indeed,
we have outlined above how the neutrino potentials may conspire to engineer significant neutrino/antineutrino transformation even when $\delta m^2$ values are small.

We have identified regions and plausible conditions in the post-bounce 
and post-shock supernova environment where the BDS, 
or something approximating it could obtain. Likewise, we have found 
that early universe scenarios with significant lepton numbers also 
provide conditions favorable for the BDS solution to reign. 
It is not yet clear that either of these venues provides a clear and 
compelling evolutionary path into the BDS regime. However, 
numerical simulations have provided hints that something like the 
BDS may occur in these environments. A central question that 
we leave for the computational neutrino flavor evolution community 
is whether, and/or to what extent, the BDS is attained. 

The stakes may be high. If neutrino energy spectra or fluxes for the 
different neutrino flavors are appreciably different at any point in 
the some $20\,{\rm s}$ time frame following core bounce, then neutrino 
and antineutrino mixing could affect shock re-heating physics, neutrino 
heating feedback on the neutrino-driven wind, and slow outflow r-process 
scenarios. Of course, the neutrino signal could be affected by any kind 
of neutrino/antineutrino flavor mixing. The effect we point out here, 
if it is ever realized deep in the supernova envelope, could appreciably 
alter the emergent neutrino energy spectra and fluxes over those 
calculated via conventional MSW evolution alone.

Finally, our considerations extend to any environment where neutrino 
fluxes are appreciable and where neutrino flavor mixing may have 
important consequences for the neutron-to-proton ratio and/or energetics 
and dynamics. Gamma-ray burst fireball models sited in the vicinity of 
a hot or collapsed compact object are a case in point. In this 
environment, the supernova problem, and in the early universe we are 
hard pressed to follow computationally the evolution of the 
neutrino/antineutrino component. It is unsatisfactory that this remains true 
even in the face of the tremendous strides in experimental neutrino 
physics which have given us the neutrino mass-squared differences and 
most of the vacuum mixing parameters. 

This work was supported in part by NSF grant PHY-00-99499, 
the TSI collaboration's DOE SciDAC grant at UCSD (G.M.F.), and
DOE grant DE-FG02-87ER40328 at UMN (Y.-Z.Q.). We thank 
P. Amanik, A. B. Balantekin, S. Bruenn, J. Carlson, H. Duan, A. Friedland, 
W. C. Haxton, W. Landry, C. Lunardini, A. Mezzacappa,
R. Sawyer, and H. Y\"uksel 
for useful discussions. We also thank the Institute for Nuclear 
Theory at the University of Washington for hospitality.

\end{document}